\documentclass[journal,10pt]{IEEEtran}

\usepackage{graphicx,amsmath,amssymb}
\usepackage{subfigure}
\usepackage{cite}
\usepackage{epstopdf}
\usepackage{fancyhdr}
\usepackage{mdwmath}
\usepackage{mdwtab}
\usepackage{balance}
\usepackage{xcolor}
\usepackage{bm}
\usepackage{amsthm}
\usepackage{algorithm}
\usepackage{multirow}
\usepackage{flafter}
\usepackage{setspace}
\usepackage{booktabs}
\usepackage{tabularx}
\usepackage{threeparttable}
\usepackage{stfloats}
\usepackage{algorithmic}
\usepackage{mathrsfs}
\usepackage{enumerate}
\usepackage{url}
\usepackage{caption}
\usepackage{lipsum}
\usepackage{gensymb}
\usepackage{multicol}

\allowdisplaybreaks[4]
\newtheorem{theorem}{Theorem}

\newtheorem{lemma}{Lemma}

\newtheorem{corollary}{Corollary}

\newtheorem{assumption}{Assumption}

\newtheorem{guideline}{Guideline}

\newtheorem{definition}{Definition}

\newtheorem{remark}{Remark}

\begin{document}
\begin{sloppypar}

\title{Multi-Objective Optimisation of URLLC-Based Metaverse Services}

\author{{Xinyu~Gao,~\IEEEmembership{Member,~IEEE,}
Wenqiang~Yi,~\IEEEmembership{Member,~IEEE}\\
Yuanwei~Liu,~\IEEEmembership{Senior Member,~IEEE}, and 
Lajos~Hanzo,~\IEEEmembership{Life Fellow,~IEEE}

\thanks{Part of this work has been invited for publication at the 18th International Symposium on Wireless Communication Systems (ISWCS), 19-22 October, 2022. (Invited paper) \cite{invitation}.}
\thanks{X. Gao, W. Yi, and Y. Liu are with the School of Electronic Engineering and Computer Science, Queen Mary University of London, London E1 4NS, U.K. (e-mail:\{x.gao,w.yi,yuanwei.liu,\}@qmul.ac.uk).}
\thanks{L. Hanzo is with the Department of Electronics and Computer Science, University of Southampton, Southampton SO17 1BJ, U.K. (e-mail:lh@ecs.soton.ac.uk).}
\thanks{L. Hanzo would like to acknowledge the financial support of the  Engineering and Physical Sciences Research Council projects EP/W016605/1, EP/X01228X/1 and EP/X01228X/1 as well as of the European Research Council's Advanced Fellow Grant QuantCom (Grant No. 789028)}
\thanks{This work was also supported in part by the CHIST-ERA grant SUNRISE CHIST-ERA-20-SICT-005, in part by the Engineering and Physical Sciences Research Council under Project EP/W035588/1, and in part by the PHC Alliance Franco-British Joint Research Programme under Grant 822326028.}
}}

\maketitle

\begin{abstract}
    Metaverse aims for building a fully immersive virtual shared space, where the users are able to engage in various activities. To successfully deploy the service for each user, the Metaverse service provider and network service provider generally localise the user first and then support the communication between the base station (BS) and the user. A reconfigurable intelligent surface (RIS) is capable of creating a reflected link between the BS and the user to enhance line-of-sight. Furthermore, the new key performance indicators (KPIs) in Metaverse, such as its energy-consumption-dependent total service cost and transmission latency, are often overlooked in ultra-reliable low latency communication (URLLC) designs, which have to be carefully considered in next-generation URLLC (xURLLC) regimes. In this paper, our design objective is to jointly optimise the transmit power, the RIS phase shifts, and the decoding error probability to simultaneously minimise the total service cost and transmission latency and approach the Pareto Front (PF). We conceive a twin-stage central controller, which aims for localising the users first and then supports the communication between the BS and users. In the first stage, we localise the Metaverse users, where the stochastic gradient descent (SGD) algorithm is invoked for accurate user localisation. In the second stage, a meta-learning-based position-dependent multi-objective soft actor and critic (MO-SAC) algorithm is proposed to approach the PF between the total service cost and transmission latency and to further optimise the latency-dependent reliability. Our numerical results demonstrate that 1) The proposed solution strikes a tradeoff between the total service cost and transmission latency, which provides a candidate group of optimal solutions for diverse practical scenarios. 2) The proposed meta-learning-based MO-SAC algorithm is capable of adaption to new wireless environments, compared to the benchmarkers. 3) The approximate PF depicted discovered the relationships among the KPIs for the Metaverse, which provides guidelines for its deployment.
\end{abstract}



\section{Introduction}
The Metaverse is constituted by a network of three-dimensional (3D) virtual worlds relying on the social connections of computer-generated applications and the physical world. It hinges on technologies that empower multisensory connections with virtual environments, digital objects, and individuals \cite{Joshua}. Although in each virtual application, the localisation and communication can be processed separately. Metaverse has different applications in the same virtual world, so the localisation and communication for different users are overlapping at the same time. Hence, to avoid interference between localisation and communication stages, different operating frequencies-based physical-layer solutions are sought to support stringent localisation and data rate requirements for each user. Compared to visible light communication or vision-based technologies, millimetre wave (mmWave) technology is less affected by natural light and atmospheric media \cite{why}. Furthermore, the locations of outer mmWave transmitters are known and they are regarded as accurate, while the localisation trackers of built-in are usually inaccurate since the errors accumulate \cite{Gao1}. To this end, mmWave techniques can be used for localising users, and the non-overlapping TeraHertz (THz) band may be harnessed for providing a Gbps-level data rate without imposing self-interference from sensing echoes. Additionally, the wireless sensing and communications services in Metaverse tend to rely on direct line-of-sight (LOS) propagation between the base station (BS) and users, but this may be blocked by obstacles \cite{HZhang}. The popular reconfigurable intelligent surfaces (RISs) \cite{CPan,QYi,Yuanwei}, are capable of creating a reflected path in a cost-efficient manner for improving both the sensing accuracy and the spectral efficiency.
\par
The 5G-style ultra-reliable low latency communication (URLLC) may be harnessed for early Metaverse applications on a small scale \cite{CShe1,CShe2,Van Huynh}, as demonstrated in extended reality (XR) scenarios \cite{XR1, XR2, XR3}. However, the evolved large-scale Metaverse imposes even more stringent localisation, data rate, and energy efficiency requirements on next-generation URLLC (xURLLC) \cite{Lu,Du}. In addition to the aforementioned physical-layer challenges of the xURLLC-enabled Metaverse, we have to comprehensively consider the quality of experience (QoE) versus energy cost trade-off.

\vspace{-0.5cm}
\subsection{State-of-the-art}


\subsubsection{Evolution of Wireless Communication Aided Metaverse}
To support the ubiquitous coverage of Metaverse, flawless wireless transmission is necessary \cite{FTang}. With the global research momentum ramping up surrounding the Metaverse, XR is one of the most tangible manifestations of the Metaverse at the current state-of-the-art. Briefly, XR is an umbrella term that covers virtual reality (VR), augmented reality (AR), and mixed reality. A novel VR model based on multi-attribute utility theory was proposed by Chen \emph{et al.} \cite{MChen} for capturing the VR users' quality of service. The echo state network-based learning algorithm \cite{Network} is proposed for collecting the tracking information gleaned from VR users, and for transmitting video and the accompanying audio to the VR users via wireless channels. As a further advance, Taha \emph{et al.} \cite{ATaha} developed a comprehensive red, green, and blue (RGB)-based framework for constructing accurate localisation and high-resolution depth maps using mmWave systems. The simulations showed the promising gains of mmWave-based depth perception compared to the conventional RGB approaches \cite{RGB} in VR scenarios. Then Batalla \emph{et al.} \cite{Jordi Mongay Batalla} analysed the quality of VR/AR video streamed over WiFi and 5G networks in harsh industrial environments. Indeed, in order to maintain low latency, while outsourcing computation in the face of mobility, the mobile edge cloud will play a key role in next-generation networks. In this context, a novel performance metric namely the XR quality index (XQI) was proposed by Dou \emph{et al.} \cite{Shengyue Dou} to reflect the impact of networking-induced imperfections on XR services. Specifically, both fine-grained and coarse-grained XQI models are provided for producing a final score that can reflect the impact of realistic imperfect networking on the QoE. The Metaverse relies on the elements of XR and combines them with the conventional Internet. To elaborate, there are several emerging considerations in the Metaverse. For example, to achieve customised high-quality Metaverse services, a user-attention-aware network resource alposition was designed by Du \emph{et al.} \cite{HDU}. To minimise the communication load, a sampling, communication, and prediction co-design framework was proposed by Meng \emph{et al.} \cite{ZMeng} subject to a constraint on tracking the mean squared error between a real-world device and its digital model in the Metaverse. To minimise the cost to the service provider, Ng \emph{et al.} \cite{Ng} conceived a virtual education case study in the Metaverse and solved the associated unified resource alposition problem in the face of stochastic user demand. With the development of the Metaverse, the range of key performance indicators (KPIs) has also been extended to the associated wireless network services. The roll-out of large-scale applications imposes high costs on the Metaverse service providers (MSP) and network service providers (NSP). Therefore, the energy-related total operational expenses, namely total service cost, should also be considered as a KPI in Metaverse.

\subsubsection{Evolution from URLLC to xURLLC}
In this context, Xie \emph{et al.} \cite{HailiangXie} studied the URLLC downlink, where a RIS assisted a BS sends individual short-packet messages to multiple users. They solved the associated latency minimisation problem via alternating optimisation. Then the resource alposition of RIS-assisted multiple-input single-output orthogonal frequency division multiple access (OFDMA)-aided multicell networks was investigated by Ghanem \emph{et al.} \cite{WalidRGhanem}, where a set of cooperating BSs served a set of URLLC users. As a further development, Hashemi \emph{et al.} \cite{RaminHashemi} proposed a multi-objective (MO) optimisation problem for maximising the achievable finite blocklength rate, while minimising the blocklengths of a RIS-assisted system. In addition to the latency and reliability requirements considered in traditional URLLC, Haber \emph{et al.} \cite{ElieElHaber} optimised the use of the RIS along with the design of the offloading and resource alposition parameters for maximising the sum rate of the users while satisfying stringent reliability specifications. In \cite{JingxuanZhang}, the self-adaptive flexible transmission time interval scheduling strategy of the enhanced mobile broadband and URLLC coexistence scenario was proposed for improving the reliability of the system. Then Chaccour \emph{et al.} \cite{Christina Chaccour} investigated whether the data-rate QoE requirements can be met in case of wireless connectivity in the THz frequency bands. Ren \emph{et al.} \cite{HRen} proposed to deploy RISs for enhancing the transmission reliability under specific data rate constraints. A RIS-assisted wireless communication system relying on non-linear energy harvesting and ultra-reliable low-latency constraints was designed by Dohk \emph{et al.} \cite{Dhok} for applications in industrial automation. Furthermore, a URLLC-aided relaying system was considered in \cite{Ranjha}, where the design objective was to minimise the total energy subject to optimal resource alposition.

\vspace{-0.5cm}
\subsection{Motivations and Contributions}
The xURLLC enhances real-time interactions, immersive experiences, and collaboration within the Metaverse by providing ultra-reliable and low-latency communication. Although the existing literature outlined the benefits of the communication-aided Metaverse and the foundations of xURLLC, there are still numerous crucial research challenges. Firstly, multiple KPIs, such as the total service cost and transmission latency, have to be optimised simultaneously in the emerging xURLLC-enabled Metaverse. The conflicting relationships among these KPIs typically result in a MO optimisation problem. Secondly, accurate interference-free localisation and communication are of salient significance in this context. Thirdly, improving the channel conditions by RISs based on the specific positions of the users is still a challenge. We are inspired to solve these problems, hence the main contributions of this work can be summarised as follows:
\begin{itemize}
    \item We propose a self-interference-free twin-stage central controller for an indoor RIS-assisted Metaverse scenario, which intrinsically amalgamates mmWave and non-overlapping THZ for localising the users and supports communication between the BS and users. To analyse the performance of the proposed network in the Metaverse system, new KPIs, such as the total service cost as well as transmission latency are defined and considered. Since the bit-level reliability attained and the transmission latency in the physical layer are correlated, our goal is to strike a balance between the total service cost and the transmission latency of the Metaverse users.
    \item We design a stochastic gradient descent (SGD)-based algorithm for approaching the best-case localisation performance characterised by the Cram\'er-Rao method. We adopt a localisation-dependent meta-learning-based MO-soft-actor-and-critic (SAC) algorithm to train a policy, which aims for adapting to new tasks, such as new wireless environments. The core point of this strategy is to integrate the action values acquired by random sampling preferences and assign dynamic weights for both the loss functions for further updating the policy by homotopy optimisation. We demonstrate that this algorithm is capable of adapting to the next tasks in the face of uncertainty. 
    \item Our proposed solution strikes a tradeoff between the total service cost and transmission latency, which provides a candidate group of near-optimal solutions for practical scenarios. The approximate Pareto Front (PF) portrays the relationships among the KPIs for the Metaverse, which provides guidelines for the application's deployment for MSP and NSP. Compared to conventional MO-SAC and single-objective algorithms, our meta-learning-based MO-SAC algorithm adapts more promptly to new tasks upon increasing the number of training samples and approaches the PF.
\end{itemize}

\vspace{-0.5cm}
\subsection{Orgainisations}
The rest of this paper is organised as follows. Section II presents our system model followed by the formulation of our simultaneous total service cost and latency optimisation problems. In Section III, we propose a bespoke meta-learning-based MO-SAC algorithm. Section IV presents our numerical results for quantifying the meta-learning-based MO-SAC algorithm's performance. Finally, Section V concludes this paper.



\vspace{-0.5cm}
\section{System Model and Problem Formulation}
\begin{figure*}[htbp]
    \centering  
    \setlength{\belowcaptionskip}{-0.6cm}
    \includegraphics[scale = 0.14]{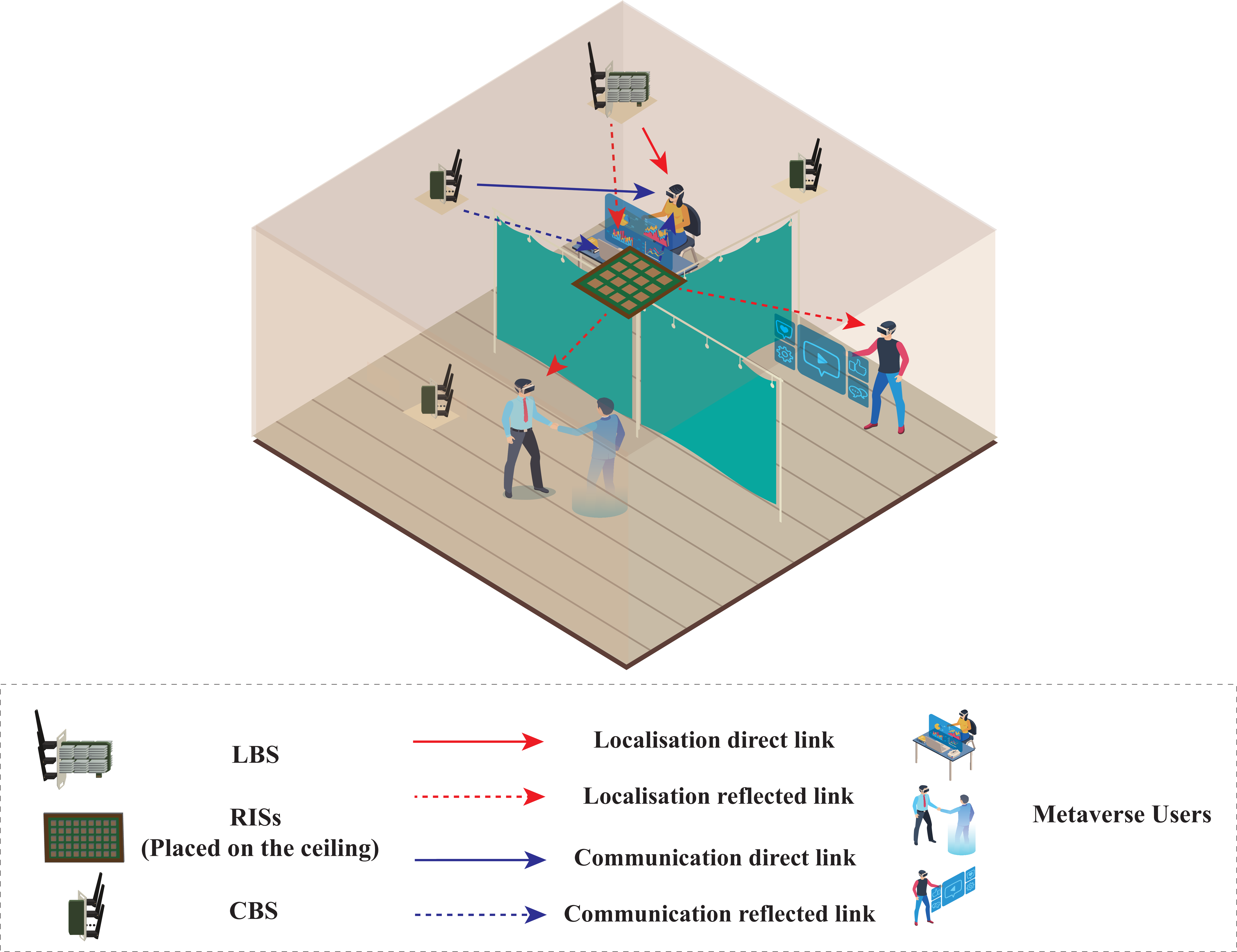}
    \caption{Illustration of an Metaverse scenario considered.}
    \label{VR}
\end{figure*}

In this paper, we focus our attention on an indoor Metaverse scenario consisting of $U$ single-antenna Metaverse users. To support Metaverse services for the users, we consider a twin-stage central controller to assist their actions. The \textit{first stage} determines the position of Metaverse users by a mmWave localisation base station (LBS), namely a mmWave Wi-Fi router, having $N$ antennas, where the RIS is able to create a reliable reflected link between the LBS and the blocked Metaverse users. The \textit{second stage} supports information communication between the $B$ THz communication base stations (CBSs) and the Metaverse users, where the RIS enhances the link quality by mitigating the attenuation between the CBS and Metaverse users. The RIS equipped with $K = BK_b$ reconfigurable elements is employed for assisting the hybrid wireless network, where $B$ and $K_b$ denote the number of sub-surfaces and the number of elements in each sub-surface, respectively \footnote{Note that to take full advantage of the RIS resources, the number of RIS sub-surfaces is consistent with the number of CBSs, while each CBS is able to build a link with the sub-surface.}. The CBSs and LBS are distributed on the wall, while the RIS is located on the ceiling \footnote{We assume that the vertical height ratio of the ceiling and Metaverse users is high enough so that the RIS is able to locate all the Metaverse users without blockage in its fields of view.}. Due to the narrow pencil-beams of THz systems \cite{HZhang}, each CBS only serves a single user at each time slot $t$, i.e. $U \leq B$. In this model, we focus on the communication performance within a time period $\mathcal{T}$, which consists of $T$ time slots. Then, the aforementioned two stages will be introduced in more depth in the following subsections.

\vspace{-0.5cm}
\subsection{Localisation Stage}
In this stage, the LBS and RIS cooperate to locate the Metaverse users. Let us assume having a 3D Cartesian coordinate system, where the origin is at the centre of the floor, as shown in Fig.~\ref{VR}. The positions of the LBS, the $u$-th Metaverse user, and the RIS are given by $\mathbf{p}_L$ = ($x_L, y_L, h_L$), $\mathbf{p}_{u,t}$ = ($x_{u,t}, y_{u,t}, h_{u,t}$), and $\mathbf{p}_r$ = ($x_r, y_r, h_r$), respectively. The LBS transmits a continuous mmWave wave OFDM signal in a resource block containing $M$ adjacent subcarriers. Since the frequencies reach the mmWave and Terahertz band, the NLOS components of the channel models can be ignored. Hence, the path loss, and AOA and AOD of the LOS component can be determined, once the BS and RIS are deployed, which can be further regarded as deterministic. The channel gain between the LBS and $u$-th Metaverse user at the $m$-th subcarrier at time slot $t$, $t \in [0, T]$ can be expressed as:
\par
\vspace{-0.1cm}
\noindent
\begin{align}\label{localization total channel}
    (\mathbf{h}_{u,t}^{m})^{H} = (\mathbf{h}_{\mathrm{L},u,t}^{m})^{H} + (\mathbf{h}_{\mathrm{R},u,t}^{m})^{H}\pmb{\Theta}(t)\mathbf{H}_{\mathrm{L},\mathrm{R},t}^{m},
\end{align}
\par
\vspace{-0.1cm}
\noindent
where $\pmb{\Theta}$(t) = diag($e^{j\theta_{1},t}, e^{j\theta_{2},t}, \cdots, e^{j\theta_{K},t}$) is the reflection coefficient matrix of RIS. Accordingly, the receive signal can be expresses as:
\par
\vspace{-0.1cm}
\noindent
\begin{align}\label{localization total signal}
    Y_{u,t}^{m} = (\mathbf{h}_{u,t}^{m})^{H}X_{u,t}^{m} + n_0,
\end{align}
\par
\vspace{-0.1cm}
\noindent
where variable $n_0$ is the circularly symmetric complex Gaussian noise with zero mean and variance $\nu^2$. The channels spanning from the LBS to the $u$-th Metaverse user $(\mathbf{h}_{\mathrm{L},u,t}^{m})^{H} \in \mathbb{C}^{1 \times N}$, from the RIS to $u$-th Metaverse user $(\mathbf{h}_{\mathrm{R},u,t}^{m})^{H} \in \mathbb{C}^{1 \times K}$, and from LBS to RIS $\mathbf{H}_{\mathrm{L},\mathrm{R},t}^{m} \in \mathbb{C}^{K \times N}$ are defined as \footnote{The elements in RIS are deployed edge-to-edge in the isotropic scattering environment considered, and an omnidirectional antenna LBS is deployed \cite{35,36}.}:
\par
\vspace{-0.1cm}
\noindent
\begin{align}\label{localization BU channel}
    &(\mathbf{h}_{\mathrm{L},u,t}^{m})^{H} = \alpha_{\mathrm{L},u,t}e^{-j2\pi m\frac{\tau_{\mathrm{L},u,t}}{M\mathcal{B}}}\mathbf{a}_{\mathrm{L}}^{H}(\omega_{\mathrm{L},u,t}), \\
    &(\mathbf{h}_{\mathrm{R},u,t}^{m})^{H} = \alpha_{\mathrm{R},u,t}e^{-j2\pi m\frac{\tau_{\mathrm{R},u,t}}{M\mathcal{B}}}\mathbf{a}_{\mathrm{R}}^{H}(\omega_{\mathrm{R},u,t}), \\
    &\mathbf{H}_{\mathrm{L},\mathrm{R},t}^{m} = \alpha_{\mathrm{L},\mathrm{R},t}e^{-j2\pi m\frac{\tau_{\mathrm{L},\mathrm{R},t}}{M\mathcal{B}}}\mathbf{a}_{\mathrm{R}}(\varphi_{\mathrm{L},\mathrm{R},t})\mathbf{a}_{\mathrm{L}}^{H}(\omega_{\mathrm{L},\mathrm{R},t}), 
\end{align}
\par
\vspace{-0.1cm}
\noindent
where $\alpha_{c_0,c_1,t} = \rho_{c_0,c_1,t}e^{j\phi_{c_0,c_1,t}}$ has the modulus $\rho_{c_0,c_1,t}$ and phase $\phi_{c_0,c_1,t}, c_0 \in \{\mathrm{L},\mathrm{R}\}, c_1 \in \{\mathrm{R},u\}$. The variables $\varphi_{c_0,c_1,t}$, $\omega_{c_0,c_1,t}$, and $\tau_{c_0,c_1,t} = ||\hat{p}_{c_0,c_1,t}||/c + \eta_0$ are the angle-of-arrival (AOA), the angle-of-departure (AOD), and the delay. The variables $\hat{p}_{c_0,c_1,t}$, $c$, and $\eta_0$ are the distance between $c_0$ and $c_1$, speed of light, and clock offset, respectively. The variable $\mathcal{B} = 1/B_0$ represents the sampling time with $B_0$ being the bandwidth. The transmit steering vector (TSV) in the localisation stage is given by:
\par
\vspace{-0.1cm}
\noindent
\begin{align}\label{ASR}
    &\mathbf{a}_{c_0}^{H}(\omega_{c_0,c_1,t}) = \nonumber \\
    &\hspace{3em} [1, e^{j\frac{2\pi}{\lambda_{c}}z\sin\omega_{c_0,c_1,t}}, \cdots, e^{j(N-1)\frac{2\pi}{\lambda_{c}}z\sin\omega_{c_0,c_1,t}}]^{H},
\end{align}
\par
\vspace{-0.1cm}
\noindent
where $\lambda_{c}$ and $z$ are the mmWave wavelength and the antenna element spacing, respectively. Let us denote the estimated position error and the accurate position area by $\mathbf{p}^e_{u,t}$ = ($x^e_{u,t}, y^e_{u,t}, h^e_{u,t}$), and $\mathbf{p}^*_{u,t}$ = ($x^*_{u,t}, y^*_{u,t}, h^*_{u,t}$), respectively. The estimated positions of $u$-th user at time slot $t$ can be expressed as:
\par
\vspace{-0.1cm}
\noindent
\begin{align}\label{localization state}
    \mathbf{p}_{u,t} \in [\mathbf{p}^*_{u,t} - \mathbf{p}^e_{u,t}, \mathbf{p}^*_{u,t} + \mathbf{p}^e_{u,t}], \hspace{1em} \eta_0 < \overline{\eta},
\end{align}
\par
\vspace{-0.1cm}
\noindent
where $\overline{\eta}$ is the maximum tolerable clock offset determining the maximum synchronisation error for localisation.

\vspace{-0.5cm}

\subsection{Communication Stage}
In the second stage, the central controller pairs each CBS with one of the Metaverse users based on the localisation results. Each CBS orientates its main lobe direction towards the associated user. Let $s_{b,u,t} \in \{0,1\}$ denote the index of the link status between the $b$-th CBS and the $u$-th user at time slot $t$. If $s_{b,u,t} = 1$, the link is established; otherwise, $s_{b,u,t} = 0$. The transmit gain of each CBS and the receive gain of each user is given by \cite{CLin}:
\par
\vspace{-0.1cm}
\noindent
\begin{align}\label{transmit gain}
    G_{\mathrm{Tr}} = \frac{4\pi}{\delta_{\mathrm{Tr}} + 1}\Omega_{\mathrm{Tr}},
\end{align}
\begin{align}\label{receive gain}
    G_{\mathrm{Re}} = \frac{4\pi}{\delta_{\mathrm{Re}} + 1}\Omega_{\mathrm{Re}},
\end{align}
\par
\vspace{-0.1cm}
\noindent
where $\delta_{(\beta)}, \beta\in\{\mathrm{Tr}, \mathrm{Re}\}$ is the power ratio between the side lobes and the main lobe of the transmit antenna (TA) and receive antenna (RA). We have $\Omega_{\beta}$ $ = 4\arcsin(\tan(\psi_{H}^{\beta}/2)\tan(\psi_{V}^{\beta}/2))$, where $\psi_{H}^{\beta}$ and $\psi_{V}^{\beta}$ denote the horizontal and vertical beam widths of the TA and RA, respectively. Let $\hat{g}_{b,b,u,t} \in \mathbb{C}^{1 \times K_b}$ represent the cascaded channel gain from the $b$-th CBS to the $u$-th Metaverse user via the $b$-th sub-surface. Thus, the cascaded channel from the $b$-th CBS to the $u$-th link via the $b$-th sub-surface in the RIS is given by \cite{Boulogeorgos}:
\par
\vspace{-0.1cm}
\noindent
\begin{align}\label{CBS-RIS-user gain}
    g_{b,b,u,t}(s_{b,u,t}) = s_{b,u,t}\hat{g}_{b, b, u, t}\mathbf{a}_{\mathrm{R}}(\omega_{b,u,f,t})\pmb{\Theta}(t)\mathbf{a}_{\mathrm{S}}^{H}(\varphi_{b,b,f,t}),
\end{align}
\par
\vspace{-0.1cm}
\noindent
where $\omega_{b,u,f,t}$, $\varphi_{b,b,f,t}$ are the AOA from $b$-th sub-surface to $u$-th user and AOD from $b$-th CBS to the $b$-th sub-surface, respectively. The TSVs from the $b$-th sub-surface of the RIS to the $u$-th user and from the $b$-th CBS to the $b$-th sub-surface of the RIS can be rewritten as:
\par
\vspace{-0.1cm}
\noindent
\begin{align}\label{TSVs}
    &\mathbf{a}_{\mathrm{R}}(\omega_{b,u,f,t}) \nonumber \\
    &\hspace{1em} = [e^{-j\omega_{b_1,u,f,t}},\cdots,e^{-j\omega_{b_k,u,f,t}},\cdots,e^{-j\omega_{b_{K_b},u,f,t}}], \\
    &\mathbf{a}_{\mathrm{S}}^{H}(\varphi_{b,b,f,t}) \nonumber \\
    &\hspace{1em} = [e^{-j\varphi_{b,b_1,f,t}},\cdots,e^{-j\varphi_{b,b_k,f,t}},\cdots,e^{-j\varphi_{b,b_{K_b},f,t}}],
\end{align}
\par
\vspace{-0.1cm}
\noindent
where the variables $\omega_{b_k,u,f,t}$ and $\varphi_{b,b_k,f,t}$ are the phase differences of the incoming signal to the $k$-th reflecting element relative to the first element $b_1$ of the $b$-th sub-surface, and the signal to the $u$-th user reflected from the $k$-th reflecting element relative to the first element $b_1$ of the $b$-th sub-surface, respectively. According to \cite{YPan}, the variable $\hat{g}_{b, b, u, t}$ in \eqref{CBS-RIS-user gain} can be expressed as:
\par
\vspace{-0.1cm}
\noindent
\begin{align}\label{CBS-RIS-user gain without phase}
    \hat{g}_{b, b, u, t} = \frac{\sqrt{G_{\mathrm{Tr}}G_{\mathrm{Re}}}c}{8\sqrt{\pi^{3}}f\sigma_{u,b}}e^{-j\frac{2\pi f}{c}d_{b,b,u,t}}e^{-\frac{1}{2}V(f)d_{b,b,u,t}},
\end{align}
\par
\vspace{-0.1cm}
\noindent
where $\sigma_{u,b}$, and $f$ are the reference distance of the $b$-th sub-surface, and the THz operating frequency, respectively. Furthermore, $d_{b,b,u,t} = |[x_{b_1} - x_{u,t}, y_{b_1} - y_{u,t}, z_{b_1} - z_{u,t}]| + |[x_b - x_{b_1}, y_b - y_{b_1}, z_b - z_{b_1}]|$, where ($x_{b_1}, y_{b_1}, z_{b, 1}$) is the coordinate of the first element of the $b$-th sub-surface of the RIS. Finally, $e^{-V(f)}$ denotes the medium transmittance obeying Beer-Lambert law with $V(f)$, which is regarded as the overall absorption medium coefficient at frequency $f$ \cite{VPetrov}. The path loss of the direct link between $b$-th SBS and $u$-th user can be given by \cite{Christina Chaccour}:
\par
\vspace{-0.1cm}
\noindent
\begin{align}\label{path loss}
    &\mathcal{L}_{b,u,t}(s_{b,u,t}) = \nonumber \\
    & \hspace{2em} \left\{
        \begin{array}{lr}
            \frac{\sqrt{G_{\mathrm{Tr}}G_{\mathrm{Re}}}c}{8\sqrt{\pi^{3}}f\sigma_{u,b}}e^{-\frac{1}{2}V(f)E_{b,u,t}(s_{b,u,t})}, \hspace{0.5em} s_{b,u,t} = 1, & \\
          \hspace{5.5em} 0, \hspace{6.75em} s_{b,u,t} = 0,
        \end{array}
    \right.
\end{align}
\par
\vspace{-0.1cm}
\noindent
where $E_{b,u,t}(s_{b,u,t})$ denotes the distance between $b$-th SBS and $u$-th user.

\subsubsection{Data Rate}
In the system studied, the achievable data rate is defined as the ratio of the number of information bits to the number of symbols transmitted. The signal-to-noise ratio (SNR) $\varrho$ of the $b$-th CBS to the $u$-th user via the $b$-th sub-surface of the RIS at time slot $t$ can be expressed as:
\par
\vspace{-0.1cm}
\noindent
\begin{align}\label{SNR}
    \varrho_{b,b,u,t}(s_{b,u,t}) = \frac{P_t|g_{b,b,u,t}(s_{b,u,t}) + \mathcal{L}_{b,u,t}(s_{b,u,t})|^2}{n^2_{u,t}(s_{b,u,t})},
\end{align}
\par
\vspace{-0.1cm}
\noindent
where $n^2_{u,t}(s_{b,u,t})$ denotes the noise power. Hence the associated achievable data rate can be expressed as follows \cite{WLiu}:
\par
\vspace{-0.1cm}
\noindent
\begin{align}\label{data rate}
    &D_{b,b,u,t}(s_{b,u,t}) = \nonumber \\
    &\hspace{1em} s_{b,u,t}\Bigg(\mathrm{log}_2\bigg(1+\varrho_{b,b,u,t}(s_{b,u,t})\bigg) - \sqrt{\frac{\hat{D}}{m}}\frac{Q^{-1}(\varepsilon)}{\ln2}\Bigg),
\end{align}
\par
\vspace{-0.1cm}
\noindent
where $m$ and $\varepsilon$ are the transmission block length and the decoding error probability (DEP). $Q^{-1}(x)$ is the inverse of the function $Q(x) = 1/\sqrt{2\pi}\int^{\infty}_{x}e^{-t^2/2}dt$, while $\hat{D} = 1-\big(1+\varrho_{b,b,u,t}(s_{b,u,t})\big)^2$.

\subsubsection{Transmission Latency}
Assuming that the transmitted data size assigned to the $u$-th user at time slot $t$ is $S_{u,t}$, the resultant transmission delay can be formulated as:
\par
\vspace{-0.1cm}
\noindent
\begin{align}\label{delay}
    J_{u,t}(s_{u,t}) = \frac{S_{u,t}}{\sum_{b=1}^{B}D_{b,b,u,t}(s_{b,u,t})}.
\end{align}
\par
\vspace{-0.1cm}
\noindent
Then, the transmission state of the $u$-th user at time slot $t$ can be expressed as:
\par
\vspace{-0.1cm}
\noindent
\begin{align}\label{delay state}
    \mathcal{K}_{u,t}(\mathbf{s}_{u,t}) = \left\{
        \begin{array}{lr}
         1, \hspace{0.5em} J_{u,t}(\mathbf{s}_{u,t}) \leq \Delta t, & \\
         0, \hspace{0.5em} \mathrm{Otherwise},
        \end{array}
      \right.
\end{align}
\par
\vspace{-0.1cm}
\noindent
where $\Delta t$ is the maximum tolerable transmission delay, and $\mathbf{s}_{u,t} = [s_{1,u,t},s_{2,u,t},\cdots,s_{B,u,t}]$. Then we define a index $\overline{J}_{u,t}(s_{u,t})$ to represent the maximum transmission latency among Metaverse users at each time slot, which can be expressed as:
\par
\vspace{-0.1cm}
\noindent
\begin{align}\label{max delay state}
    \overline{J}_{u,t}(s_{u,t}) = \max \{J_{1,t}(s_{1,t}),\cdots,J_{u,t}(s_{u,t}),\cdots,J_{U,t}(s_{U,t})\}
\end{align}
\par
\vspace{-0.1cm}
\noindent
The latency performance of the networks considered is evaluated by transmission latency index in the networks considered.

\subsubsection{Total Service Cost}
Given its demanding specifications, the total service cost is a pivotal KPI for the Metaverse, which depends on the energy and DEP. The total service cost of the Metaverse services is determined by both the MSP and NSP, and it can be defined as follows \cite{Ng}:
\par
\vspace{-0.1cm}
\noindent
\begin{align}
    \mathcal{E}(P_t,\varepsilon) &= -C_{\mathrm{meta}} + C_{\mathrm{net}}(P_t,\varepsilon), \nonumber \\
    &= -(F_{\mathrm{meta}}^{s}+F_{\mathrm{QoE}}^{s}) + (f_P^s\sum_{t=1}^{T}P_t+f_\varepsilon^s\varepsilon)
\end{align}
\par
\vspace{-0.1cm}
\noindent
where $C_{\mathrm{meta}}$ and $C_{\mathrm{net}}(P_t,\varepsilon)$ are the cost and revenue of the entire Metaverse system, respectively. Explicitly, $C_{\mathrm{meta}}$ is determined by the fixed cost $F_{\mathrm{meta}}^{s}$ of hardware and the cost $F_{\mathrm{QoE}}^{s}$. of satisfying the QoE required. Furthermore, the $C_{\mathrm{net}}(P_t,\varepsilon)$ is determined by the cost of transmit power $f_P^s\sum_{t=1}^{T}P_t$ and the DEP $f_\varepsilon^s\varepsilon$, which has to be optimised, where $f_P^s$ and $f_\varepsilon^s$ are the fee for unit $P_t$ value and $\varepsilon$ value, respectively.

\vspace{-0.5cm}
\subsection{Problem Formulation}
Both the latency and the cost are important for Metaverse applications. For example, low latency ensures a smooth user experience and real-time interactions, while optimising the costs allows for scalability and accessibility, fostering wider adoption of the Metaverse. Naturally, our preference is to achieve high reliability at a low total service cost. In the model studied, the reliability and transmission latency are correlated, where a low transmission rate indicates high reliability. Therefore, our goal is to strike a balance between the total service cost and the minimal transmission latency of the Metaverse users. The minimisation problem can be formulated as follows:
\par
\vspace{-0.1cm}
\noindent
\begin{align}
    \min_{s_{b,u,t},\varepsilon,\eta_0,P_t,\pmb{\Theta}(t)} \hspace*{1em}& \bigg[\sum_{t = 1}^{T}\mathcal{E}(P_t,\varepsilon), \overline{J}_{u,t}(s_{u,t})\bigg] \label{problem}\\
    {\rm s.t.} \hspace*{1em}
    &0 < P_t \leq P_{\mathrm{max}}, \tag{\ref{problem}{a}} \label{problema}\\
    &0\le\theta_k\le 2\pi, k \in \{1,2,\cdots,K\}, \tag{\ref{problem}{b}} \label{problemb} \\
    &\sum_{b=1}^{B}s_{b,u,t}=1, \forall u \in {1,2,\cdots,U}, \tag{\ref{problem}{c}} \label{problemc} \\
    &\sum_{u=1}^{U}s_{b,u,t}=1, \forall b \in {1,2,\cdots,B}, \tag{\ref{problem}{d}} \label{problemd} \\
    &s_{b,u,t} \in \{0,1\}, \tag{\ref{problem}{e}} \label{probleme} \\
    &\varepsilon \leq \varepsilon_m, \tag{\ref{problem}{f}} \label{problemf} \\
    &\eta_0 < \overline{\eta}, \tag{\ref{problem}{g}} \label{problemg}
\end{align}
\par
\vspace{-0.1cm}
\noindent
where $P_{\mathrm{max}}$ and $\varepsilon_m$ are the maximum transmit power of the Metaverse users and the maximum tolerable DEP, respectively. Constraint \eqref{problema} limits the transmit power, while constraint \eqref{problemb} limits the phase of the RIS reflection coefficients. Furthermore, constraints \eqref{problemc}, \eqref{problemd}, and \eqref{probleme} represent to the one-to-one mapping of the CBSs and Metaverse users at time slot $t$. Constraints \eqref{problemf} and \eqref{problemg} represent the limit of $\varepsilon$ and of the maximum synchronisation error $\eta_0$ for localisation, respectively. However, problem \eqref{problem} is difficult to solve for the following four reasons. Firstly, the Metaverse users' positions change as time elapses, making the channel hard to model. Secondly, the problem formulated is considered a Markov Decision Process (MDP) problem. Thus, the requirements of any immediately consecutive time slots are governed by MDP, which does not lend itself to employment in our infinite variable optimisation. Thirdly, the total service cost of the MSP and the transmission latency of the NSP are subject to a trade-off, when optimising the DEP and transmit power. Explicitly, upon increasing the DEP and the transmit power, the total service cost increases, while the latency decreases. Hence, jointly optimising the total service cost and latency for obtaining a group of Pareto-optimal (PO) results provides us with a range of different system settings for NSP according to the prevalent practical requirements, compared to conventional methods. Fourthly, the model trained is required to achieve prompt convergence to the new tasks, but this issue is beyond the scope of conventional MO optimisation algorithms. Therefore, the traditional optimisation and reinforcement learning (RL) algorithms are unsuitable for solving the problem \eqref{problem}. As a remedy, we propose a novel meta-learning-based MO optimisation algorithm.


\section{Meta Learning-based Position-dependent MO-RL Algorithm}
\begin{figure}[htbp]
    \centering 
    \setlength{\belowcaptionskip}{-0.8cm}
    \includegraphics[scale = 0.35]{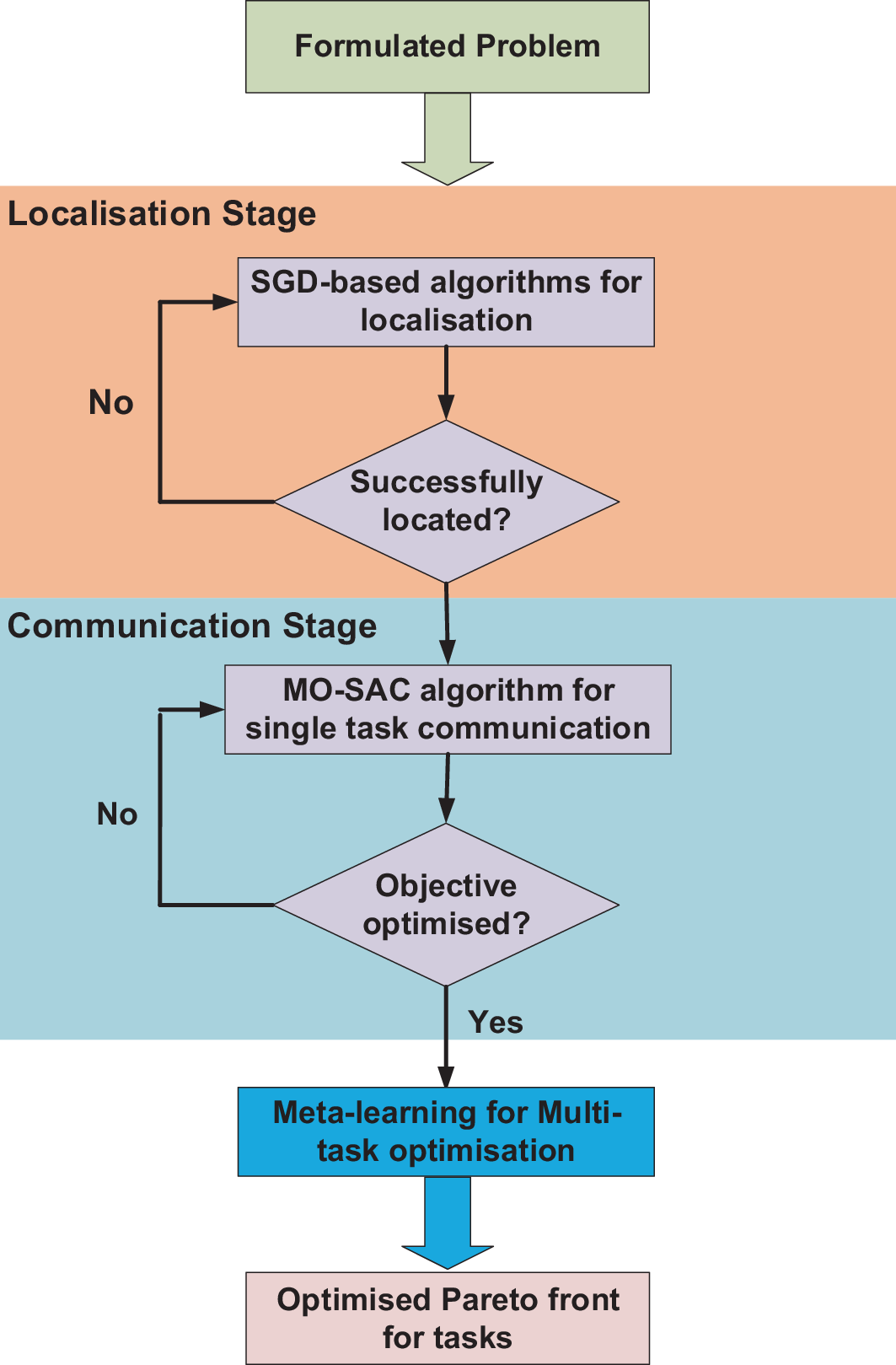}
    \caption{The sequence of the algorithms proposed for solving the problem formulated.}
    \label{timeline}
\end{figure}

In this section, we introduce a position-dependent MO-RL algorithm based on the meta-learning framework \cite{Finn}. There are two stages of the proposed algorithm: the position-dependent MO-RL learning stage, and the meta-learning stage. The position-dependent MO-RL stage is able to simultaneously optimise multiple objectives based on the estimated positions of users, while the meta-learning aims for training a model for prompt adaptation to new tasks. The timeline of proposed algorithms is shown in Fig.~\ref{timeline}.

\subsection{Position-dependent MO-RL Learning Stage}
The localisation-dependent MO-RL Learning stage aims for finding a policy to solve problem \eqref{problem}. For localisation, the Cram\'er-Rao method \cite{Cramer} is employed for calculating the lower bounds of the Metaverse users' localisation, and a policy gradient-based algorithm is invoked for estimating the positions. Then, \eqref{problem} can be optimised according to the estimated positions.

\subsubsection{Cram\'er-Rao Lower Bounds}
The steps are briefly presented for highlighting the lower bounds of the localisation estimation of the user position $\mathbf{p}_{u,t}$ and clock offset $\eta_0$. According to the channel model of the localisation stage, the channel parameter vector of $u$-th user at time slot $t$ can be expressed as follows:
\par
\vspace{-0.1cm}
\noindent
\begin{align}\label{localization channel vector}
    &\overline{\mathbf{q}}_{u, t} = [\tau_{\mathrm{L},u,t}, \omega_{\mathrm{L},u,t}, \rho_{\mathrm{L},u,t}, \phi_{\mathrm{L},u,t}, \tau_{\mathrm{L},u,t}, \omega_{\mathrm{L},u,t}, \nonumber \\ 
    &\hspace{8.5em} (\rho_{\mathrm{L},\mathrm{R},t}\rho_{\mathrm{R},u,t}), (\phi_{\mathrm{L},\mathrm{R},t} + \phi_{\mathrm{R},u,t})].
\end{align}
\par
\vspace{-0.1cm}
\noindent
Upon denoting the unbiased estimator $\overline{\mathbf{q}}^{e}_{u, t}$ of $\overline{\mathbf{q}}_{u, t}$, the corresponding Fisher information matrix (FIM) $\mathcal{J}_{\overline{\mathbf{q}}_{u, t}} \in \mathbb{C}^{8 \times 8}$ should satisfing \cite{Cramer}:
\par
\vspace{-0.1cm}
\noindent
\begin{align}\label{Cramer bounds}
    \mathbb{E}\{(\overline{\mathbf{q}}^{e}_{u, t}-\overline{\mathbf{q}}_{u, t})(\overline{\mathbf{q}}^{e}_{u, t}-\overline{\mathbf{q}}_{u, t})^H\} \succeq \mathcal{J}_{\overline{\mathbf{q}}_{u, t}}^{-1},
\end{align}
\par
\vspace{-0.1cm}
\noindent
where $\mathbf{E}_0 \succeq \mathbf{F}_0$ means that $\mathbf{E}_0 - \mathbf{F}_0$ is positive semi-definite. Accordingly, the $(a, b)$-th FIM entry $[\mathcal{J}_{\overline{\mathbf{q}}_{u, t}}]_{(a,b)}$ can be expressed according to \textbf{Lemma \ref{lemma 1}}:
\begin{align}\label{localization channel FIM}
    [\mathcal{J}_{\overline{\mathbf{q}}_{u, t}}]_{(a,b)} = \frac{2}{\epsilon^2}\sum_{m=0}^{M-1}\mathcal{R}\Big\{\Big(\frac{\partial \tilde{h}^m_{u,t}}{\overline{q}^a_{u, t}}\Big)^H\frac{\partial \tilde{h}^m_{u,t}}{\overline{q}^b_{u, t}}\Big\},
\end{align}
where we have $\tilde{h}^m_{u,t} = [\sqrt{P_0} (\mathbf{h}_{\mathrm{L},u,t}^{m}+ (\mathbf{h}_{\mathrm{R},u,t}^{m})^{H} \pmb{\Theta}(t) \mathbf{H}_{\mathrm{LBS},\mathrm{R},t}^{m})^{H}]X^m_{u,t}$. The variables $P_0$ and $X^m_{u,t}$ represent the constant transmit power of mmWave communication and transmitted symbol of the $m$-th subcarrier at time slot $t$. 

\begin{lemma}\label{lemma 1}
    For a vector $\mathbf{Z}$ which follows a symmetric complex Gaussian distribution ($Z_0,Z_1$), the $(a, b)$-th entry of its Fisher information matrix is given by the Slepian-Bangs formula [44] of:
    \begin{align}
        [\mathcal{J}_{\mathbf{Z}}]_{(a,b)} = 2\mathcal{R}\Big\{\Big(\frac{\partial Z_0}{\mathbf{Z}^a}\Big)^HZ_1^{-1}\frac{\partial Z_0}{\mathbf{Z}^b}\Big\} + \mathcal{T}\Big\{Z_1^{-1}\frac{Z_1}{\mathbf{Z}^a}Z_1^{-1}\frac{Z_1}{\mathbf{Z}^b}\Big\}.
    \end{align}
\end{lemma}

\noindent
Therefore, the FIM can be expressed in the channel domain as follows:
\par
\vspace{-0.1cm}
\noindent
\begin{align}\label{FIM in channel domain}
    \mathcal{J}_{\overline{\mathbf{q}}_{u, t}} = 
        \begin{bmatrix}
        \mathcal{J}_{\overline{\mathbf{q}}_{\mathrm{L},u,t}} & \mathcal{J}_{c} \\
        \mathcal{J}_{c} & \mathcal{J}_{\overline{\mathbf{q}}_{\mathrm{L},\mathrm{R},u,t}} 
        \end{bmatrix},
\end{align}
\par
\vspace{-0.1cm}
\noindent
where $\mathcal{J}_{\overline{\mathbf{q}}_{\mathrm{L},u,t}} \in \mathbb{C}^{4 \times 4}$ and $\mathcal{J}_{\overline{\mathbf{q}}_{\mathrm{L},\mathrm{R},u,t}} \in \mathbb{C}^{4 \times 4}$ are the FIM submatrices corresponding to the LOS path and the non-line-of-sight (NLOS) path, respectively. The variable $\mathcal{J}_{c}$ is the LOS-NLoS path cross-correlation. To obtain the FIM in the position domain, the position parameter vector $\hat{\mathbf{q}}_{u, t}$ can be transformed from the unknown channel parameters vector $\overline{\mathbf{q}}_{u, t}$ as follows:
\par
\vspace{-0.1cm}
\noindent
\begin{align}\label{localization vector}
    &\hat{\mathbf{q}}_{u, t} = [x_{u,t}, y_{u,t}, h_{u,t}, \rho_{\mathrm{L},u,t}, \phi_{\mathrm{L},u,t}, \nonumber \\
    &\hspace{7.5em}(\rho_{\mathrm{L},\mathrm{R},t}\rho_{\mathrm{R},u,t}), (\phi_{\mathrm{L},\mathrm{R},t} + \phi_{\mathrm{R},u,t}), \eta_0],
\end{align}
\par
\vspace{-0.1cm}
\noindent
and the corresponding FIM in the position domain is given by:
\par
\vspace{-0.1cm}
\noindent
\begin{align}\label{localization FIM}
    \mathcal{J}_{\hat{\mathbf{q}}_{u, t}} = \frac{\partial \overline{\mathbf{q}}_{u, t}^H}{\hat{\mathbf{q}}_{u, t}}\mathcal{J}_{\overline{\mathbf{q}}_{u, t}}\Big(\frac{\partial \overline{\mathbf{q}}_{u, t}^H}{\hat{\mathbf{q}}_{u, t}}\Big)^{H}.
\end{align}
\par
\vspace{-0.1cm}
\noindent
Therefore, the lower bounds of the $u$-th user's position estimate and the clock offset estimate can be derived from the diagonal elements of the corresponding inverted FIM $\mathcal{J}_{\hat{\mathbf{q}}_{u, t}}$. The position error bound (PEB) of the $u$-th user at time slot $t$ is formulated as:
\par
\vspace{-0.1cm}
\noindent
\begin{align}\label{PEB1}
    \mathrm{PEB}_{u, t} = \sqrt{[\mathcal{J}_{\hat{\mathbf{q}}_{u, t}}]_{(1,1)}^{-1}+[\mathcal{J}_{\hat{\mathbf{q}}_{u, t}}]_{(2,2)}^{-1}+[\mathcal{J}_{\hat{\mathbf{q}}_{u, t}}]_{(3,3)}^{-1}} \hspace{0.5em}.
\end{align}
\par
\vspace{-0.1cm}
\noindent
Then equation \eqref{localization state} can be written as:
\par
\vspace{-0.1cm}
\noindent
\begin{align}\label{localization state1}
    \mathbf{p}_{u,t} \in [(\mathbf{p}^*_{u,t} - \mathrm{PEB}_{u, t}), (\mathbf{p}^*_{u,t} + \mathrm{PEB}_{u, t})], \hspace{0.5em} \eta_0 = \eta^e_0,
\end{align}
\par
\vspace{-0.1cm}
\noindent
where $\eta^e_0$ is the estimated clock offset.

\subsubsection{SGD-based Algorithm for Position Estimation}
The Cram\'er-Rao bound is important for the Metaverse as it provides a benchmark for assessing the achievable accuracy in parameter estimation. This is relevant for tracking user movements, locations, or gestures, helping optimise the reliability of sensor-based applications within the Metaverse. However, this method is the general solution to evaluate the positions, our goal is to minimise the lower bound to approach the accurate estimations. Based on the lower bounds, we design the direct estimator $\overline{\mathbf{p}}_{u,t} = [x_{u,t}, y_{u,t}, h_{u,t}, \eta]$ of $u$-th Metaverse user at time slot $t$. The signal $\mathbf{Y}_{u,t}$ received over $M$ subcarriers can be expressed as:
\par
\vspace{-0.1cm}
\noindent
\begin{align}\label{Localization Signal}
    \mathbf{Y}_{u,t} = \sqrt{P_0}\pmb{\iota}\pmb{\alpha} + n_{0},
\end{align} 
\par
\vspace{-0.1cm}
\noindent
where we have $\pmb{\alpha} = [\alpha_{\mathrm{L},u,t}, (\alpha_{\mathrm{L},\mathrm{R},t}\alpha_{\mathrm{R},u,t})]$, and $\pmb{\iota} = [\mathbf{X}_{\mathrm{L},u,t}^{H}\mathbf{a}_{\mathrm{L}}(\omega_{\mathrm{L},u,t}),\mathbf{X}_{\mathrm{L},\mathrm{R},u,t}^{H}(\mathbf{a}_{\mathrm{R}}(\varphi_{\mathrm{L},\mathrm{R},t})\mathbf{a}_{\mathrm{L}}^{H}(\omega_{\mathrm{L},\mathrm{R},t}))^H\pmb{\Theta}(t)$ $\mathbf{a}_{\mathrm{R}}^{H}(\omega_{\mathrm{R},u,t})]$. The $\mathbf{X}_{\mathrm{L},u,t} = [X_{\mathrm{L},u,t}^{0}, e^{-j2\pi /M\mathcal{B}}X_{\mathrm{L},u,t}^{1},$ $\cdots, e^{-j2\pi (M-1)/M\mathcal{B}}X_{\mathrm{L},u,t}^{M-1}]$, while $\pmb{\alpha}$ is the unknown parameter vector, then the estimation problem can be formulated as
\par
\vspace{-0.1cm}
\noindent
\begin{align}\label{localization estimation}
    \overline{\mathbf{p}}_{u,t}^{*} = \mathrm{arg} \min_{\overline{\mathbf{p}}_{u,t}}[\min_{\pmb{\alpha}} \mathcal{L}(\overline{\mathbf{p}}_{u,t}, \pmb{\alpha})],
\end{align}
\par
\vspace{-0.1cm}
\noindent
where $\mathcal{L}(\overline{\mathbf{p}}_{u,t}, \pmb{\alpha}) = ||\mathbf{Y}_{u,t} - \sqrt{P_0}\pmb{\iota}\pmb{\alpha}||^{2}$. This likelihood function can be minimised with respect to $\pmb{\alpha}$, which can be rewritten as
\par
\vspace{-0.1cm}
\noindent
\begin{align}\label{d}
    \overline{\mathcal{L}}(\overline{\mathbf{p}}_{u,t}) = ||\mathbf{Y}_{u,t} - \sqrt{P_0}\pmb{\iota}(\overline{\mathbf{p}}_{u,t})\hat{\pmb{\alpha}}(\overline{\mathbf{p}}_{u,t})||^{2},
\end{align}
\par
\vspace{-0.1cm}
\noindent
where $\hat{\pmb{\alpha}} = \frac{1}{\sqrt{P_0}}(\pmb{\iota}^{H}\pmb{\iota})^{-1}(\pmb{\iota}^{H})\mathbf{Y}_{u,t}$. The policy gradient can be written as:
\par
\vspace{-0.1cm}
\noindent
\begin{align}\label{dloss}
    \overline{\gamma} &= \overline{\gamma} + \frac{1}{\hat{\mathcal{B}}}\nabla_{\overline{\mathbf{p}}_{u,t}} \overline{\mathcal{L}}^{\hat{b}}(\overline{\mathbf{p}}_{u,t}), \nonumber \\
    &= \overline{\gamma} + \frac{1}{\hat{\mathcal{B}}}\nabla_{\overline{\mathbf{p}}_{u,t}}||\mathbf{Y}_{u,t}^{\hat{b}} - \sqrt{P_0}\pmb{\iota}^{\hat{b}}(\overline{\mathbf{p}}_{u,t})\hat{\pmb{\alpha}}^{\hat{b}}(\overline{\mathbf{p}}_{u,t})||^{2},
\end{align}
\par
\vspace{-0.1cm}
\noindent
where $\overline{\gamma}$, $\hat{\mathcal{B}}$, and $\hat{b}$ are the parameters of the SGD algorithm, the minibatch of training data, and the $\hat{b}$-th training sample, respectively. Therefore, the final optimised estimator $\overline{\mathbf{p}}_{u,t}$ for the $u$-th user at time slot $t$ can be obtained according to the SGD algorithm. The pseudo-code of the SGD algorithm is provided in \textbf{Algorithm~\ref{Localization algorithm}}.

\begin{algorithm}
    \caption{SGD Algorithm with minibatch on Localisation}
    \label{Localization algorithm}
    \begin{algorithmic}[1]
    \REQUIRE ~~\\ 
    Number of epochs for SGD algorithm $\mathcal{V}$, mini-batch size for SGD algorithm $\hat{\mathcal{B}}$, learning rate $\hat{\eta}$, an initial parameters of the SGD algorithm $\overline{\gamma}$.
    \ENSURE Optimised estimator $\overline{\mathbf{p}}_{u,t}$.
    \STATE Initial positions error $\mathbf{p}^{\mathrm{ini}}_{u,t}$ for $u$-th user at time slot $t$ according to the PEB.
    \FOR {epoch = 1, 2, $\cdots$, $\mathcal{V}$}
        \IF {stopping criterion not met}
            \WHILE {stopping criterion not met}
                \STATE Random sample $\hat{\mathcal{B}}$ examples from the normalised training set.
                \STATE Initialise $\overline{\gamma}$.
                \FOR {iteration = 1, 2, $\cdots$, $\hat{\mathcal{B}}$}
                    \STATE Compute policy gradient: \\
                    $\overline{\gamma} = \overline{\gamma} + \frac{1}{\hat{\mathcal{B}}}\nabla_{\overline{\mathbf{p}}_{u,t}}||\mathbf{Y}_{u,t}^{\hat{b}} - \sqrt{P_0}\pmb{\iota}^{\hat{b}}(\overline{\mathbf{p}}_{u,t})\hat{\pmb{\alpha}}^{\hat{b}}(\overline{\mathbf{p}}_{u,t})||^{2}$.
                \ENDFOR
                \STATE Update $\overline{\gamma} \leftarrow \overline{\gamma} - \hat{\eta}\overline{\gamma}$.
            \ENDWHILE
        \ENDIF
    \ENDFOR
    \STATE Obtain the optimised positions error $\mathbf{p}^e_{u,t}$ from $\mathcal{V}$ iterations.
    \STATE Calculate the estimated position $\mathbf{p}_{u,t} \in [\mathbf{p}^*_{u,t} - \mathbf{p}^e_{u,t}, \mathbf{p}^*_{u,t} + \mathbf{p}^e_{u,t}]$.
    \end{algorithmic}
  \end{algorithm}

\subsubsection{MO-RL Learning Stage}

Based on the positions, a MO-based SAC algorithm is invoked \cite{SAC}. In SAC, the MDP can be written as $\langle \overline{\mathbf{S}}, \overline{\mathbf{A}}, \mathbf{p}, \overline{\mathbf{R}},\gamma_0\rangle$, where $\overline{\mathbf{S}}$, $\overline{\mathbf{A}}$ and $\gamma_0$ denote the state space, action space, and the discount factor, respectively. The unknown transition probability $\mathbf{p}: \overline{\mathbf{S}} \times \overline{\mathbf{S}} \times \overline{\mathbf{A}} \rightarrow [0, \infty)$ represents the probability of a transition from the current state $\overline{\mathbf{s}}_t \in \overline{\mathbf{S}}$ to the next state $\overline{\mathbf{s}}_{t+1} \in \overline{\mathbf{S}}$, while the $\overline{\mathbf{r}}$ is the reward vector bounded $\overline{\mathbf{r}}: \overline{\mathbf{S}} \times \overline{\mathbf{A}} \rightarrow [\overline{r}_{\min}, \overline{r}_{\max}]$. In the system studied, the controller acts as the agent, while the variables $\overline{\mathbf{s}}_t$, $\overline{\mathbf{a}}_t$, and $\overline{\mathbf{r}}$ can be expressed as follows: $\overline{\mathbf{s}}_t = [\theta_t, P_t]$, $\overline{\mathbf{a}}_t = [\Delta \theta, \Delta P]$, $\overline{\mathbf{r}} = [\Delta (\sum_{b = 1}^{B} \mathcal{E}(P_{t+1},\varepsilon)^b - \sum_{b = 1}^{B} \mathcal{E}(P_t,\varepsilon)^b), \Delta (J_{u,t+1}(s_{u,t+1}) - J_{u,t}(s_{u,t}))]$. The variables $\zeta_{\pi}(\overline{\mathbf{s}}_t)$ and $\zeta_{\pi}(\overline{\mathbf{s}}_t, \overline{\mathbf{a}}_t)$ denote the trajectories of the state and the state-action marginals under the policy $\pi(\overline{\mathbf{a}}_t|\overline{\mathbf{s}}_t)$. A general maximum-entropy objective having the expected entropy over $\zeta_{\pi (\overline{\mathbf{s}}_t)}$ can be expressed as \cite{SAC}:
\par
\vspace{-0.1cm}
\noindent
\begin{align}\label{Expected entropy}
    \varpi(\pi) = \sum_{t=0}^{T}\mathbb{E}_{(\overline{\mathbf{s}}_t, \overline{\mathbf{a}}_t) \sim \zeta_{\pi}}[\overline{\mathbf{r}}(\overline{\mathbf{s}}_t, \overline{\mathbf{a}}_t) + \epsilon\mathcal{H}(\pi(\cdot|\overline{\mathbf{s}}_t))],
\end{align}
\par
\vspace{-0.1cm}
\noindent
where $\epsilon$ is the temperature parameter. To extend \eqref{Expected entropy} to the MO scenario, a linear preference vector $\mathbf{w} \subseteq \mathbf{W}$ is invoked for estimating the expected total rewards. The goal of preferences is to assign specific weights for the objectives, where the importance of each objective remains the same under each preference. The \eqref{Expected entropy} can be rewritten as:
\par
\vspace{-0.1cm}
\noindent
\begin{align}\label{MO Expected entropy}
    \varpi(\pi, \mathbf{w}) = \sum_{t=0}^{T}\mathbb{E}_{(\overline{\mathbf{s}}_t, \overline{\mathbf{a}}_t, \mathbf{w}) \sim \zeta_{\pi}}[\overline{\mathbf{r}}(\overline{\mathbf{s}}_t, \overline{\mathbf{a}}_t, \mathbf{w}) + \epsilon\mathcal{H}(\pi(\cdot|\overline{\mathbf{s}}_t), \mathbf{w})].
\end{align}
\par
\vspace{-0.1cm}
\noindent
The parameter $\epsilon$ controls the grade of stochasticity of the optimal policy by determining the relative importance of the entropy $\mathcal{H}[\pi(\cdot|\overline{\mathbf{s}}_t, \mathbf{w})]$. However, since $\epsilon$ is subsumed into the reward by scaling it according to $\epsilon^{-1}$, it can be omitted in the rest of the paper. As for the fixed policy, the soft Q-value can be iteratively computed and repeated by applying the Bellman backup operator $\mathcal{D}^{\pi}$, which can be formulated as
\par
\vspace{-0.1cm}
\noindent
\begin{align}\label{MOQ}
    \mathcal{D}^{\pi}\mathbf{Q}(\overline{\mathbf{s}}_t, \overline{\mathbf{a}}_t, \mathbf{w}) = \overline{\mathbf{r}}(\overline{\mathbf{s}}_t, \overline{\mathbf{a}}_t, \mathbf{w}) + \gamma_0\mathbb{E}_{\overline{\mathbf{s}}_{t+1} \sim \zeta_{\pi}}[V(\overline{\mathbf{s}}_{t+1}, \mathbf{w})],
\end{align}
\par
\vspace{-0.1cm}
\noindent
where $V(\overline{\mathbf{s}}_{t+1}, \mathbf{w}) = \mathbb{E}_{\overline{\mathbf{a}}_{t+1} \sim \pi}[\mathbf{Q}(\overline{\mathbf{s}}_t, \overline{\mathbf{a}}_t,\mathbf{w}) - \mathrm{log}\pi(\overline{\mathbf{a}}_t|\overline{\mathbf{s}}_t, \mathbf{w})]$ is the soft state value function. Then, we define an optimality filter $\mathcal{G}$ for \eqref{MOQ} as follows:
\par
\vspace{-0.1cm}
\noindent
\begin{align}\label{argQ}
    \mathcal{G}\mathbf{Q}(\overline{\mathbf{s}}_t, \mathbf{w}) = \mathrm{arg}_{\mathbf{Q}} \mathrm{sup}_{\overline{\mathbf{a}}_t \in \overline{\mathbf{A}}, \mathbf{w}' \subseteq \mathbf{W}} \mathbf{w}^H \mathbf{Q}(\overline{\mathbf{s}}_t, \overline{\mathbf{a}}_t, \mathbf{w}'),
\end{align}
\par
\vspace{-0.1cm}
\noindent
where $\mathrm{arg}_Q$ represents the maximum Q value operation and $\mathrm{sup}(\cdot)$ is the supremum operation. The mo optimality operator $\mathcal{D}$ may then defined as:
\par
\vspace{-0.1cm}
\noindent
\begin{align}\label{MOQ optimal}
    &\mathcal{D}\mathbf{Q}(\overline{\mathbf{s}}_t, \overline{\mathbf{a}}_t, \mathbf{w}) = \overline{\mathbf{r}}(\overline{\mathbf{s}}_t, \overline{\mathbf{a}}_t, \mathbf{w}) + \nonumber \\
    & \gamma_0\mathbb{E}_{\overline{\mathbf{s}}_{t+1} \sim \zeta_{\pi}}\{\mathbb{E}_{\overline{\mathbf{a}}_{t+1} \sim \pi}[\mathcal{G}\mathbf{Q}(\overline{\mathbf{s}}_t, \overline{\mathbf{a}}_t,\mathbf{w}) - \mathrm{log}\pi(\overline{\mathbf{a}}_t|\overline{\mathbf{s}}_t, \mathbf{w})]\}.
\end{align}
\par
\vspace{-0.1cm}
\noindent
\textbf{Theorems \ref{theorem 1} - \ref{theorem 3}} demonstrate the feasibility of the mo optimality operator $\mathcal{D}$ of our MO-RL.

\begin{theorem}\label{theorem 1}
    Let $\mathbf{Q}'$ be the preferred optimal value function as follows:
    \begin{align}\label{pargQp}
        &\mathbf{Q}'(\overline{\mathbf{s}}_t, \overline{\mathbf{a}}_t, \mathbf{w}) = \Big\{\mathrm{arg}_{\mathbf{Q}} \mathrm{sup}_{\pi \in \Pi} \mathbf{w}^H \cdot \nonumber \\
        &\mathbb{E}_{\overline{\mathbf{s}}_{t+1} \sim \zeta_{\pi}}  \Big[\sum_{t=0}^{T} \gamma_0^{t} \overline{\mathbf{r}}(\overline{\mathbf{s}}_t, \overline{\mathbf{a}}_t, \mathbf{w})\Big] - \mathbf{w}^{'H}\mathrm{log}\pi(\overline{\mathbf{a}}_t|\overline{\mathbf{s}}_t, \mathbf{w})\Big\} .
    \end{align}
    Then, it can be shown that: $\mathbf{V}'(\overline{\mathbf{s}}_t, \overline{\mathbf{a}}_t, \mathbf{w}) = \mathcal{G} \mathbf{Q}'(\overline{\mathbf{s}}_t, \overline{\mathbf{a}}_t, \mathbf{w})$.
    \par
    \vspace{-0.1cm}
    \noindent
    \begin{proof}
        See Appendix~A.
    \end{proof}
    \par
    \vspace{-0.2cm}
    \noindent
\end{theorem}

\begin{theorem}\label{theorem 2}
    Let us denote $\mathbf{Q}$ and $\hat{\mathbf{Q}}$ as any two MO Q value functions. The Lipschitz condition $\overline{d}(\mathcal{G}\mathbf{Q}, \mathcal{G}\hat{\mathbf{Q}}) \leq \gamma_0 \overline{d}(\mathbf{Q}, \hat{\mathbf{Q}})$ can be obtained, where $\gamma_0$ is the discount factor.
    \par
    \vspace{-0.1cm}
    \noindent
    \begin{proof}
        See Appendix~B.
    \end{proof}
    \par
    \vspace{-0.2cm}
    \noindent
\end{theorem}

\begin{theorem}\label{theorem 3}
    If $\mathcal{G}$ can be contracted by the discount factor $\gamma_0$ across the complete pseudo-metric space $\langle\pmb{\mathcal{Q}}, \overline{d}\rangle$, it can be shown that $\mathrm{lim}_{t \rightarrow \infty} \overline{d}(\mathcal{G}^{t}, \mathbf{Q}, \hat{\mathbf{Q}}) = 0$, where $\pmb{\mathcal{Q}}$ is the value space.
    \par
    \vspace{-0.1cm}
    \noindent
    \begin{proof}
        See Appendix~C.
    \end{proof}
    \par
    \vspace{-0.2cm}
    \noindent
\end{theorem}

As mentioned above, the state value is capable of approximating the soft value. The soft value function is trained to minimise the squared residual error, which can be expressed as:
\begin{align}\label{squared residual error}
    &\varpi_{V}(\mu_0) = \mathbb{E}_{\overline{\mathbf{s}}_t \sim \mathcal{F}} \Big[\frac{1}{2}(V(\overline{\mathbf{s}}_{t}, \mathbf{w}) - \nonumber \\
    &\hspace{2em} \mathbb{E}_{\overline{\mathbf{a}}_{t} \sim \pi_{\mu_1}}[\mathbf{Q}_{\mu_2}(\overline{\mathbf{s}}_t, \overline{\mathbf{a}}_t, \mathbf{w}) - \mathrm{log}\pi_{\mu_1}(\overline{\mathbf{a}}_t|\overline{\mathbf{s}}_t, \mathbf{w})])^{2}\Big],
\end{align}
where $\mathcal{F}$ represents the previous sample states and the actions distribution or the replay buffer. Therefore, the gradient can be estimated as:
\begin{align}\label{gradient1}
    &\hat{\nabla}_{\mu_0} \varpi_{V}(\mu_0) = \nabla_{\mu_0} V_{\mu_0}(\overline{\mathbf{s}}_t)[V_{\mu_0}(\overline{\mathbf{s}}_t) - \nonumber \\
    &\hspace{7em} \mathbf{Q}_{\mu_2}(\overline{\mathbf{s}}_t, \overline{\mathbf{a}}_t, \mathbf{w}) + \mathrm{log}\pi_{\mu_1}(\overline{\mathbf{a}}_t|\overline{\mathbf{s}}_t, \mathbf{w})],
\end{align}
where the actions are sampled according to the current policy. The soft Bellman residual can be calculated by:
\begin{align}\label{Bellman residual}
    &\varpi_{\mathbf{Q}}(\mu_2) = \mathbb{E}_{(\overline{\mathbf{s}}_t,\overline{\mathbf{a}}_t) \sim \mathcal{F}}\big[ \frac{1}{2}(\mathbf{Q}_{\mu_2}(\overline{\mathbf{s}}_t, \overline{\mathbf{a}}_t, \mathbf{w}) - \nonumber \\
    &\hspace{4em} \overline{\mathbf{R}}(\overline{\mathbf{s}}_t, \overline{\mathbf{a}}_t, \mathbf{w}) - \gamma_0\mathbb{E}_{\overline{\mathbf{s}}_{t+1} \sim \zeta_{\pi}}V_{\hat{\mu}_0}(\overline{\mathbf{s}}_{t+1}, \mathbf{w}))^2 \big],
\end{align}
and the stochastic gradients can be expressed as:
\begin{align}\label{gradient2}
    &\hat{\nabla}_{\mu_2}\varpi_{\mathbf{Q}}(\mu_2) = \nabla_{\mu_2}\mathbf{Q}_{\mu_2}(\overline{\mathbf{s}}_t, \overline{\mathbf{a}}_t, \mathbf{w})[(\mathbf{Q}_{\mu_2}(\overline{\mathbf{s}}_t, \overline{\mathbf{a}}_t, \mathbf{w}) - \nonumber \\
    &\hspace{7.5em} \overline{\mathbf{R}}(\overline{\mathbf{s}}_t, \overline{\mathbf{a}}_t, \mathbf{w}) - \gamma_0V_{\mu_0}(\overline{\mathbf{s}}_{t+1}, \mathbf{w}))].
\end{align}
The update policy method can be expressed as:
\begin{align}\label{update policy}
    &\varpi_{\pi}(\mu_1) = \mathbb{E}_{\overline{\mathbf{s}}_t \sim \mathcal{F}, \epsilon_t \sim \mathcal{N}}[\mathbf{log}\pi_{\mu_1}(\overline{f}_{\mu_1}(\epsilon_t;\overline{\mathbf{s}}_t)|\overline{\mathbf{s}}_t,\mathbf{w}) - \nonumber \\
    &\hspace{11em} \mathbf{Q}_{\mu_2}(\overline{\mathbf{s}}_t, \overline{f}_{\mu_1}(\epsilon_t;\overline{\mathbf{s}}_t), \mathbf{w})],
\end{align}
and the approximate gradient can be formulated as:
\begin{align}\label{gradient3}
    &\hat{\nabla}_{\mu_1}\varpi_{\pi}(\mu_1) = \nabla_{\mu_1}\mathrm{log}\pi(\overline{\mathbf{a}}_t|\overline{\mathbf{s}}_t, \mathbf{w}) + [\nabla_{\overline{\mathbf{a}}_t}\mathrm{log}\pi(\overline{\mathbf{a}}_t|\overline{\mathbf{s}}_t, \mathbf{w}) - \nonumber \\
    &\hspace{8.5em} \nabla_{\overline{\mathbf{a}}_t}\mathbf{Q}(\overline{\mathbf{s}}_t, \overline{\mathbf{a}}_t, \mathbf{w})]\nabla_{\mu_1}\overline{f}_{\mu_1}(\epsilon_t;\overline{\mathbf{s}}_t),
\end{align}
where $\epsilon_t$ and $\overline{f}_{\mu_1}(\epsilon_t;\overline{\mathbf{s}}_t)$ denote an input noise vector and a neural network transformation, respectively. By optimising these gradients, we aim for approaching the PF. Therefore, the associated gradient optimisation can be regarded as multiple gradient optimisation constructed for all objectives. According to the Karush-Kuhn-Tucker (KKT) conditions, there exists a set of $\nu_1,\nu_2,\cdots,\nu_M$ for $M$ objectives so that:
\begin{itemize}
  \item  $\nu_1,\nu_2,\cdots,\nu_M \geq 0$.
  \item  $\sum_{m=1}^{M}\nu_m = 1$.
  \item  $\sum_{m=1}^{M}\nu_m \hat{\nabla}_{\mu_0} \varpi_{V}^{m}(\mu_0) = 0$, $\sum_{m=1}^{M}\nu_m \hat{\nabla}_{\mu_2} \varpi_{\mathbf{Q}}^{m}(\mu_2) = 0$, $\sum_{m=1}^{M}\nu_m \hat{\nabla}_{\mu_1}\varpi_{\pi}^m(\mu_1) = 0$.
\end{itemize}
\noindent
Since the objectives may have values of the different scales and multiple gradient optimisation is sensitive to the different ranges, the range of gradient function has been limited to [0, 1].

\begin{definition}\label{definition 1}
    A solution $\mu_{(\overline{u})}^1$ dominates a solution $\mu_{(\overline{u})}^2, \overline{u} = \{0, 1, 2\}$ if for all objectives they satisfying: $\hat{\nabla}_{\mu_{0}^1} \varpi_{V}^{m}(\mu_{0}^1) \leq \hat{\nabla}_{\mu_{0}^2} \varpi_{V}^{m}(\mu_{0}^2)$, $\hat{\nabla}_{\mu_{2}^1} \varpi_{Q}^{m}(\mu_{2}^1) \leq \hat{\nabla}_{\mu_{2}^2} \varpi_{Q}^{m}(\mu_{2}^2)$, $\hat{\nabla}_{\mu_{1}^1} \varpi_{\pi}^{m}(\mu_{1}^1) \leq \hat{\nabla}_{\mu_{1}^2} \varpi_{\pi}^{m}(\mu_{1}^2)$, while there is at least one objective satisfying $\hat{\nabla}_{\mu_{0}^1} \varpi_{V}^{n}(\mu_{0}^1) \leq \hat{\nabla}_{\mu_{0}^2} \varpi_{V}^{n}(\mu_{0}^2)$, $\hat{\nabla}_{\mu_{2}^1} \varpi_{Q}^{n}(\mu_{2}^1) \leq \hat{\nabla}_{\mu_{2}^2} \varpi_{Q}^{n}(\mu_{2}^2)$, $\hat{\nabla}_{\mu_{1}^1} \varpi_{\pi}^{n}(\mu_{1}^1) \leq \hat{\nabla}_{\mu_{1}^2} \varpi_{\pi}^{n}(\mu_{1}^2)$, $\forall m,n \in \{1,2,\cdots,M\}$.
\end{definition}

\begin{definition}\label{definition 2}
    A solution $\mu_{(\overline{u})}^1$ constitutes a PO solution, if no other solution $\mu_{(\overline{u})}^2$ dominates $\mu_{(\overline{u})}^2, \overline{u} = \{0, 1, 2\}$.
\end{definition}

\begin{definition}\label{definition 3}
    The full set of non-dominated solutions constitutes a PO set.
\end{definition}

The solution that satisfies the \textbf{definitions \ref{definition 1} - \ref{definition 3}} is defined as a PO solution. Thus, the optimisation problem can be formulated as follows:
\begin{align}\label{QCOP}
  \min_{\nu_1^0,\cdots,\nu_M^0} \Big\{ \Big|\Big|\sum_{m=1}^{M}\nu_m^0 \hat{\nabla}_{\mu_0} \varpi_{V}^{m}(\mu_0)\Big|\Big|^2_2, \Big| \sum_{m=1}^{M}\nu_m^0 = 1, \nu_m^0 \geq 0 \Big\}, \nonumber \\
  \min_{\nu_1^1,\cdots,\nu_M^1} \Big\{\Big|\Big|\sum_{m=1}^{M}\nu_m^1 \hat{\nabla}_{\mu^1}\varpi_{\pi}^m(\mu_1)\Big|\Big|^2_2 \Big| \sum_{m=1}^{M}\nu^1_m = 1, \nu^1_m \geq 0 \Big\}, \nonumber \\
  \min_{\nu_1^2,\cdots,\nu_M^2} \Big\{ \Big|\Big|\sum_{m=1}^{M}\nu_m^2 \hat{\nabla}_{\mu_2} \varpi_{\mathbf{Q}}^{m}(\mu_2)\Big|\Big|^2_2 \Big| \sum_{m=1}^{M}\nu_m^2 = 1, \nu_m^2 \geq 0 \Big\},
\end{align}
\par
\noindent
where $||\cdot||^2_2$ and $\nabla_{(\cdot)}$ denote the L2 norm and gradient descent (GD) operator. Since it has two objectives in problem \eqref{problem}, Equation \eqref{QCOP} can be simplified as:
\begin{align}\label{QCOP2}
    \min_{\nu^0 \in [0,1]} ||\nu^0 \hat{\nabla}_{\mu_0} \varpi_{V}^{1}(\mu_0) + (1-\nu^0) \hat{\nabla}_{\mu_0} \varpi_{V}^{2}(\mu_0)||^2_2, \nonumber \\
    \min_{\nu^1 \in [0,1]} ||\nu^1 \hat{\nabla}_{\mu_1} \varpi_{\pi}^{1}(\mu_1) + (1-\nu^1) \hat{\nabla}_{\mu_1} \varpi_{\pi}^{2}(\mu_1)||^2_2, \nonumber \\
    \min_{\nu^2 \in [0,1]} ||\nu^2 \hat{\nabla}_{\mu_2} \varpi_{Q}^{1}(\mu_2) + (1-\nu^2) \hat{\nabla}_{\mu_2} \varpi_{Q}^{2}(\mu_2)||^2_2.
\end{align}
\par
The optimisation problem \eqref{QCOP2} is equivalent to finding a minimum-norm point in the convex hull, which is a convex quadratic problem subject to linear constraints. Thus, an analytical solution of Equation \eqref{QCOP2} can be formulated as:
\begin{align}\label{QCOP2-solution}
  \nu^0 = \bigg\{ \frac{[\hat{\nabla}_{\mu_0} \varpi_{V}^{2}(\mu_0) - \hat{\nabla}_{\mu_0} \varpi_{V}^{1}(\mu_0)]^T\hat{\nabla}_{\mu_0} \varpi_{V}^{2}(\mu_0)}{||\hat{\nabla}_{\mu_0} \varpi_{V}^{1}(\mu_0) - \hat{\nabla}_{\mu_0} \varpi_{V}^{2}(\mu_0)||^2_2} \bigg\}_{[0,1]}, \nonumber \\
  \nu^1 = \bigg\{ \frac{[\hat{\nabla}_{\mu_1} \varpi_{\pi}^{2}(\mu_1) - \hat{\nabla}_{\mu_1} \varpi_{\pi}^{1}(\mu_1)]^T\hat{\nabla}_{\mu_1} \varpi_{\pi}^{2}(\mu_1)}{||\hat{\nabla}_{\mu_1} \varpi_{\pi}^{1}(\mu_1) - \hat{\nabla}_{\mu_1} \varpi_{\pi}^{2}(\mu_1)||^2_2} \bigg\}_{[0,1]}, \nonumber \\
  \nu^2 = \bigg\{ \frac{[\hat{\nabla}_{\mu_2} \varpi_{Q}^{2}(\mu_2) - \hat{\nabla}_{\mu_2} \varpi_{Q}^{1}(\mu_2)]^T\hat{\nabla}_{\mu_2} \varpi_{Q}^{2}(\mu_2)}{||\hat{\nabla}_{\mu_2} \varpi_{Q}^{1}(\mu_2) - \hat{\nabla}_{\mu_2} \varpi_{Q}^{2}(\mu_2)||^2_2} \bigg\}_{[0,1]},
\end{align}
where $\{\}_{[0,1]}$ represents clipping $\nu$ to $[0,1]$. The pseudo-code of the MO-SAC algorithm is provided in \textbf{Algorithm~\ref{MO-SAC algorithm}}\footnote{The action does not need normalisation, since the action range is usually defined as [-1, 1] so its variance and mean can approach 0 and 1 to follow the normal distribution. The reward function can not be normalised, since the Bellman equation-based reinforcement learning does not allow reward subtract a non-zero constant, since that would destroy the reward function of the environment itself.}.

\begin{algorithm}[htbp]
    \caption{MO-SAC Algorithm}
    \label{MO-SAC algorithm}
    \begin{algorithmic}[1]
    \REQUIRE ~~\\
    The number of iterations for MO-RL learning phase $\mathcal{U}$, mini-batch size for MO-RL learning phase $\overline{\mathcal{B}}$, replay memory for MO-RL learning phase $\mathcal{D}$, preference space $\overline{\pmb{\Omega}}$, an exponentially moving average of the value network weights $\hat{\mu}_0$.
    \ENSURE Tuple $\{(\overline{\mathbf{s}}_t,\overline{\mathbf{a}}_t,\overline{\mathbf{r}}(\overline{\mathbf{s}}_t, \overline{\mathbf{a}}_t, \mathbf{w}),\overline{\mathbf{s}}_{t+1},\mathbf{w})\}$, $t = \{1, 2, \cdots, T\}$.\\
    \FOR {size = 1, 2, $\cdots$, $\overline{\mathcal{B}}$}
        \STATE Copy the parameters $\chi_n$ to the SAC networks.
        \FOR {Iteration = 1, 2, $\cdots$, $\mathcal{U}$}
            \STATE Sample $\mathbf{w}$ from preference space $\overline{\pmb{\Omega}}$.
            \STATE Sample $\overline{\mathbf{a}}_t$ from $\pi_{\mu_2}(\overline{\mathbf{a}}_t|\overline{\mathbf{s}}_t, \mathbf{w})$.
            \STATE Sample normalised $\overline{\mathbf{s}}_{t+1}$ from $\zeta_{\pi}(\overline{\mathbf{s}}_{t+1}|\overline{\mathbf{s}}_t, \overline{\mathbf{a}}_t,\mathbf{w})$.
            \STATE Sample from $\{(\overline{\mathbf{s}}_t,\overline{\mathbf{a}}_t,\overline{\mathbf{r}}(\overline{\mathbf{s}}_t, \overline{\mathbf{a}}_t, \mathbf{w}),\overline{\mathbf{s}}_{t+1},\mathbf{w})\}$.
        \ENDFOR
        \STATE Update $\mu_0$ $\leftarrow$ $\mu_0 - z_V\hat{\nabla}_{\mu_0} \varpi_{V}(\mu_0)$.
        \STATE Update $\mu_1$ $\leftarrow$ $\mu_1 - z_{\pi}\hat{\nabla}_{\mu_1}\varpi_{\pi}(\mu_1)$.
        \STATE Update $\mu_2$ $\leftarrow$ $\mu_2 - z_{\mathbf{Q}}\hat{\nabla}_{\mu_2}\varpi_{\mathbf{Q}}(\mu_2)$.
        \STATE Update $\hat{\mu}_0$ $\leftarrow$ $\tau_0\mu_0 + (1-\tau_0)\hat{\mu}_0$.
    \ENDFOR
    \end{algorithmic}
\end{algorithm}

\subsection{Meta-learning Stage}
In the meta-policy learning stage, the goal is to optimise the meta-policy parameters $\hat{\pmb{\chi}} = \{\hat{\mu}_{0}, \hat{\mu}_{1}, \hat{\mu}_{2}\}$. In the model studied, the different initial positions of Metaverse users can be defined as different tasks. There are two sub-stages for the meta-learning stage: the training and the adaptation sub-stages.

\subsubsection{Training Sub-stage}
Assume that there are $N$ tasks (MO-RL models) for training, and the support set as well as the query set of each task are defined as $\mathbb{A}_{n}$ and $\mathbb{B}_{n}$, $n \in {1,2,\cdots,N}$. In each training iteration, a weight vector $\mathbf{W} = {W_1, W_2, \cdots, W_N}$ is randomly sampled to create $N$ MO-RL models. In the inner-loop\footnote{In the inner loop, the agent learns task-specific network parameters by performing one gradient step on a task-specific loss.} update, the optimised gradient can be specified for the $n$-th task as:
\par
\noindent
\begin{align}\label{local meta update problem}
    \pmb{\chi}_n = \left\{
        \begin{array}{lr}
            \mu_{0,n} = \mathrm{arg} \min\limits_{\varpi_{V}} \hat{\nabla}_{\mu_{0,n}} \varpi_{V}(\mu_{0,n}, \mathbb{A}_{n}), & \\
            \mu_{1,n} = \mathrm{arg} \min\limits_{\varpi_{\pi}} \hat{\nabla}_{\mu_{1,n}} \varpi_{\pi}(\mu_{1,n}, \mathbb{A}_{n}),& \\
            \mu_{2,n} = \mathrm{arg} \min\limits_{\varpi_{Q}} \hat{\nabla}_{\mu_{2,n}} \varpi_{Q}(\mu_{2,n}, \mathbb{A}_{n}),
        \end{array}
      \right.
\end{align}
\par
\noindent
where $\mu_{0,n}$, $\mu_{0,n}$, $\mu_{0,n}$ are the network weights of squared residual error, soft Bellman residual, and update policy method for the $n$-th task. Then, the update process of meta-policy parameters $\pmb{\chi}_n$ of the $n$-th MO-RL model can be expressed as follows:
\par
\noindent
\begin{align}\label{local meta update}
    \pmb{\chi}_n^{\hat{T}+1} = \left\{
        \begin{array}{lr}
            \mu_{0,n}^{\hat{T}+1} \leftarrow \mu_{0,n}^{\hat{T}} - z_V\hat{\nabla}_{\mu_0,n} \mathrm{arg} \min\limits_{\varpi_{V}} \varpi_{V}(\mu_{0,n}, \mathbb{A}_{n}), & \\
            \mu_{1,n}^{\hat{T}+1} \leftarrow \mu_{1,n}^{\hat{T}} - z_{\pi}\hat{\nabla}_{\mu_1,n} \mathrm{arg} \min\limits_{\varpi_{\pi}} \varpi_{\pi}(\mu_{1,n}, \mathbb{A}_{n}),& \\
            \mu_{2,n}^{\hat{T}+1} \leftarrow \mu_{2,n}^{\hat{T}} - z_{\mathbf{Q}}\hat{\nabla}_{\mu_{2,n}} \mathrm{arg} \min\limits_{\varpi_{\mathbf{Q}}} \varpi_{\mathbf{Q}}(\mu_{2,n}, \mathbb{A}_{n}),
        \end{array}
    \right.
\end{align}
\par
\noindent
where $z_V$, $z_{\pi}$, and $z_{\mathbf{Q}}$ are the learning rates for the inner-loop update. Then, for the outer-loop\footnote{In the outer loop, the model parameters from before the inner loop update are updated to reduce the loss after the inner loop update on the individual tasks.} update, the optimised parameters can be expressed as follows:
\par
\noindent
\begin{align}\label{global meta update problem}
    \hat{\pmb{\chi}} = \left\{
        \begin{array}{lr}
            \hat{\mu_{0}} = \mathrm{arg} \min\limits_{\varpi_{V}} \hat{\nabla}_{\hat{\mu_{0}}} \varpi_{V}(\hat{\mu_{0}}, \mathbb{A}_{n}), & \\
            \hat{\mu_{1}} = \mathrm{arg} \min\limits_{\varpi_{\pi}} \hat{\nabla}_{\hat{\mu_{1}}} \varpi_{\pi}(\hat{\mu_{1}}, \mathbb{A}_{n}),& \\
            \hat{\mu_{2}} = \mathrm{arg} \min\limits_{\varpi_{Q}} \hat{\nabla}_{\hat{\mu_{2}}} \varpi_{Q}(\hat{\mu_{2}}, \mathbb{A}_{n}),
        \end{array}
      \right.
\end{align}
\par
\noindent
and the update process of meta-policy parameters $\hat{\pmb{\chi}}$ can be expressed as follows:
\par
\noindent
\begin{align}\label{global meta update}
    &\hat{\pmb{\chi}}^{\hat{T}+1} = \nonumber \\
    &\left\{
        \begin{array}{lr}
            \hat{\mu_{0}}^{\hat{T}+1} \leftarrow \hat{\mu}_{0}^{\hat{T}} - \sum_{n=1}^{N} \hat{z}_V\hat{\nabla}_{\mu_0} \mathrm{arg} \min\limits_{\varpi_{V}} \varpi_{V}(\mu_{0,n}, \mathbb{B}_{n}), & \\
            \hat{\mu}_{1}^{\hat{T}+1} \leftarrow \hat{\mu}_{1}^{\hat{T}} - \sum_{n=1}^{N} \hat{z}_{\pi}\hat{\nabla}_{\mu_1} \mathrm{arg} \min\limits_{\varpi_{\pi}} \varpi_{\pi}(\mu_{1,n}, \mathbb{B}_{n}),& \\
            \hat{\mu}_{2}^{\hat{T}+1} \leftarrow \hat{\mu}_{2}^{\hat{T}} - \sum_{n=1}^{N} \hat{z}_{\mathbf{Q}}\hat{\nabla}_{\mu_2} \mathrm{arg} \min\limits_{\varpi_{\mathbf{Q}}} \varpi_{\mathbf{Q}}(\mu_{2,n}, \mathbb{B}_{n}),
        \end{array}
    \right.
\end{align}
\noindent
where $\hat{z}_V$, $\hat{z}_{\pi}$, and $\hat{z}_{\mathbf{Q}}$ are the learning rates for the inner-loop update. 

\begin{algorithm}[htbp]
    \caption{Meta Learning-based MO-SAC Algorithm}
    \label{meta learning-based MO-SAC algorithm}
    \begin{algorithmic}[1]
    \REQUIRE ~~\\
    Mini-batch size for meta-training phase $\tilde{\mathcal{B}}$, number of iterations for meta-training stage $\mathcal{U}_{it}$, replay memory for meta-training stage $\mathcal{D}_t$, number of iterations for meta-adaptation stage $\mathcal{U}_a$, replay memory for meta-adaptation stage $\mathcal{D}_a$, inner-loop learning rates $(z_V, z_{\pi}, z_{\mathbf{Q}})$, outer-loop learning rates $(\hat{z}_V,\hat{z}_{\pi},\hat{z}_{\mathbf{Q}})$, number of training tasks $N$, number of adaptation tasks $N_a$.
    \ENSURE The approximate PFs for $N_a$ tasks.\\
    \leftline{\textit{\textbf{Meta Training Stage:}}}
    \STATE \textbf{Initialise:} meta-policy parameters $\hat{\pmb{\chi}}$, MO-RL parameters $\pmb{\chi}_n$ for the $n$-th training task, support set $\mathbb{A}_{n}$ and the query set $\mathbb{B}_{n}$ of each task, task set \{$\mathcal{N}$\}.
    \FOR {mini-batch size $\tilde{\mathcal{B}}$}
        \STATE Sample $N$ tasks from task set \{$\mathcal{N}$\}.
        \FOR {iterations = 1, 2, $\cdots$, $\mathcal{U}_{t}$}
            \FOR {training tasks = 1, 2, $\cdots$, $N$}
                \STATE Obtain the positions of users by \textbf{Algorithm~\ref{Localization algorithm}}.
                \STATE Determine the pairing index $s_{b,u,t}$ according to the distance according to the positions of CBSs and users.
                \STATE Generate tuple $\{(\overline{\mathbf{s}}_t,\overline{\mathbf{a}}_t,\overline{\mathbf{r}}(\overline{\mathbf{s}}_t, \overline{\mathbf{a}}_t, \mathbf{w}),\overline{\mathbf{s}}_{t+1},\mathbf{w})\}$, $t = \{1, 2, \cdots, T\}$ from \textbf{Algorithm~\ref{MO-SAC algorithm}}.
                \STATE Store tuple $\{(\overline{\mathbf{s}}_t,\overline{\mathbf{a}}_t,\overline{\mathbf{r}}(\overline{\mathbf{s}}_t, \overline{\mathbf{a}}_t, \mathbf{w}),\overline{\mathbf{s}}_{t+1},\mathbf{w})\}$, $t = \{1, 2, \cdots, T\}$ into memory $\mathcal{D}_t$.
                \STATE Sample ($\overline{\mathbf{s}}_n, \overline{\mathbf{a}}_n, \overline{\mathbf{r}}_n, \overline{\mathbf{s}}_n^{'}$) from $\mathcal{D}_t$ as $\mathbb{A}_n$.
                \FOR {iterations = 1, 2, $\cdots$, $\mathcal{U}_{it}$}
                    \STATE Update $\pmb{\chi}_n$ by invoking \eqref{local meta update}.
                \ENDFOR
                \STATE Calculate loss function of the $n$-th task by $\mathbb{B}_{n}$.
            \ENDFOR
            \STATE Update $\pmb{\chi}$ by invoking \eqref{global meta update}.
        \ENDFOR
    \ENDFOR \\
    \leftline{\textit{\textbf{Meta Adaptation Stage:}}}
    \FOR {iterations = 1, 2, $\cdots$, $\mathcal{U}_a$}
        \FOR {adaptation tasks = 1, 2, $\cdots$, $N_a$}
            \STATE Copy $\pmb{\chi}$ to the $n_a$-th task.
            \STATE Generate experience tuple according to \textbf{line 8}.
            \STATE Store $\mathcal{D}_a$ into replay memory $\mathcal{D}_a$.
            \STATE Update $\pmb{\chi}$ by invoking \eqref{new meta update}.
            \STATE Obtain approximate PF for the $n_a$-th task.
        \ENDFOR
    \ENDFOR
    \end{algorithmic}
\end{algorithm}

\subsubsection{Adaptation Sub-stage}
When the training is finished, the trained model aims for adapting to the new task based on the optimised parameters. The well-trained meta-model is able to achieve fast adaptation to the new task, where the parameters $\pmb{\chi}$ of the new task can be updated by:
\par
\vspace{-0.3cm}
\noindent
\begin{align}\label{new meta update}
    \pmb{\chi} = \left\{
        \begin{array}{lr}
            \mu_{0} \leftarrow \mu_{0} - z_V\hat{\nabla}_{\mu_0} \mathrm{arg} \min\limits_{\varpi_{V}} \varpi_{V}(\mu_{0}), & \\
            \mu_{1} \leftarrow \mu_{1} - z_{\pi}\hat{\nabla}_{\mu_1} \mathrm{arg} \min\limits_{\varpi_{\pi}} \varpi_{\pi}(\mu_{1}),& \\
            \mu_{2} \leftarrow \mu_{2} - z_{\mathbf{Q}}\hat{\nabla}_{\mu_{2}} \mathrm{arg} \min\limits_{\varpi_{\mathbf{Q}}} \varpi_{\mathbf{Q}}(\mu_{2}).
        \end{array}
    \right.
\end{align}
\noindent
When prompt adaptation of the meta-model achieves convergence, the resultant approximate PF may be depicted. The pseudo-code of the meta-learning-based MO-SAC algorithm is provided in \textbf{Algorithm~\ref{meta learning-based MO-SAC algorithm}}.

\vspace{-0.5cm}
\subsection{Optimality Analysis for Meta Learning-based MO-RL Algorithm}
To evaluate the optimality, the algorithms will be analysed into two aspects: 1) The structure of the algorithms and 2) The complexity of the algorithms.
\subsubsection{Algorithm Structures}
For the structure, we can observe that the proposed SGD-based algorithm depends on the learning rate. If the learning rate is too small, the convergence is slow, and if it is too large, it will oscillate near the minimum value. Hence, the proposed SGD-based algorithm is near-optimal. Moreover, the MO-SAC algorithm introduces the neural networks, where the optimised weights and bias are both approaching the global optimal results but they are difficult to achieve. Thus, we conclude that this algorithm is near-optimal.
\subsubsection{Algorithm Complexity}
The complexity of the proposed meta-learning-based MO-RL algorithm is dominated by the MO-RL stage and meta-learning stage. The MO-RL stage has four parameters from the SGD algorithm and five parameters from the MO-SAC algorithm, which are the size of the input data $\tilde{D}$, the number of epochs in the SGD algorithm $\mathcal{V}$, the minibatch size $\hat{\mathcal{B}}$ of the SGD algorithm, the number of objectives $\tilde{N}$, the size of perference space $\tilde{\Omega}$, the size of state space $\tilde{S}_{\tilde{n}}$ of the objective $\tilde{n}$, the size of action space $\tilde{A}_{\tilde{n}}$ of the objective $\tilde{n}$, the size of reward function $\tilde{R}$, the mini-batch size of the MO-SAC algorithm $\overline{\mathcal{B}}$, and the number of iteration $\mathcal{U}$. So the complexity of the MO-RL stage is on the order of O[$\tilde{D}\mathcal{V}\hat{\mathcal{B}}\tilde{R}\overline{\mathcal{B}}\mathcal{U}\tilde{\Omega}(\tilde{S}_{\tilde{n}}\tilde{A}_{\tilde{n}})^{\tilde{N}}$]. The compelxity of the Meta-learning stage is dominated by the MO-RL stage, the mini-batch size for meta-training stage $\tilde{\mathcal{B}}$, the number of iterations of the meta-training stage $\mathcal{U}_{it}$, the replay memory of the meta-training stage $\mathcal{D}_t$, the number of iterations for meta-adaptation stage $\mathcal{U}_a$, the replay memory for meta-adaptation stage $\mathcal{D}_a$, and the number of training tasks $N$. So the complexity of the meta-learning-based MO-RL can be expressed as O[$\tilde{D}\mathcal{V}\hat{\mathcal{B}}\tilde{R}\overline{\mathcal{B}}\mathcal{U}\tilde{\Omega}(\tilde{S}_{\tilde{n}}\tilde{A}_{\tilde{n}})^{\tilde{N}}(\tilde{\mathcal{B}}\mathcal{U}_{it}\mathcal{U}_a\mathcal{D}_t\mathcal{D}_a)^{N}$]. According to the system studied, the complexity of the meta Learning-based MO-RL algorithm is O[$3\mathcal{V}\hat{\mathcal{B}}\overline{\mathcal{B}}\mathcal{U}(|\overline{\mathbf{S}}_{\tilde{n}}||\overline{\mathbf{A}}_{\tilde{n}}|)^{2}(\tilde{\mathcal{B}}\mathcal{U}_{it}\mathcal{U}_a\mathcal{D}_t\mathcal{D}_a)^{N}$].

\vspace{-0.3cm}
\section{Numerical Results}

In this section, we investigate the performance of RIS-assisted wireless Metaverse networks. The users are assumed to be randomly and uniformly placed in a 10m $\times$ 10m rectangular area, as shown in Fig.~\ref{VR}. Since we assume that the vertical height ratio of the ceiling and the users is huge enough to avoid blockage between the RIS and the users, the heights of the ceiling and the users are set to 6m and 1.7m. Additionally, the total number of CBSs and users are chosen to be 4 and 4, while $P_{\mathrm{max}}$ and $\varepsilon_m$ are 100mW and $10^{-5}$. The noise power of each user is fixed to $-110$dBm. In this model, the definition of reliability in the THz regime refers to the average number of successfully served Metaverse users. At each time slot $t$, if $\mathcal{K}_{u,t}(s_{u,t}) = 1$, the $u$-th will be defined as a reliable user. Thus, the reliability of the $u$-th user at a time slot $t$ can be expressed as follows:
\par
\noindent
\begin{align}\label{reliability state}
    \upsilon_{u,t}(\mathbf{s}_{u,t}) = \mathcal{K}_{u,t}(s_{u,t}) \lor \mathcal{K}_{u,t-1}(s_{u,t-1}),
\end{align}
\par
\noindent
where $\lor$ denotes the logical ``or'' operators. The newly served users at time slot $t$ will be
\par
\noindent
\begin{align}\label{new reliability state}
    \overline{\upsilon}_{t}(\mathbf{s}_{u,t}) = \{u|\upsilon_{u,t}(\mathbf{s}_{u,t}) = 1, \upsilon_{u,t-1}(\mathbf{s}_{u,t-1}) = 0\},
\end{align}
\par
\noindent
then, the number of successfully served users in $T$ time slots can be expressed by:
\par
\noindent
\begin{align}\label{reliability}
    \Xi_{1:T}(\mathbf{s}_{1:T}) = \sum_{t=1}^{T}|\overline{\upsilon}_{t}(\mathbf{s}_{u,t})|.
\end{align}
\par
\noindent
The seamless experience of this user can be guaranteed, when the user is served successfully at each time slot. Therefore, the range of maximum tolerable transmission delay $\Delta t$ can be set between 0 to t and it is further determined according to different requirements of practical scenarios. In the model studied, we strike a trade-off between the total service cost and maximum transmission latency among all users of the communication system upon one time slot constraint and then determine $\Delta t$ according to the different requirements. The variable $\Delta t$ guides choosing the permitted transmission delay of the entire system and associated acceptable total services cost, and further determine the reliability of each user in a time period $\mathcal{T}$. Then, let us now discuss the performance of the proposed Meta MO-SAC algorithm. As for the performance of the localisation stage, the \textbf{Adam algorithm} \cite{Jais}, \textbf{Batch Gradient descent (BGD) algorithm} \cite{Ruder}, and \textbf{SGD algorithm} \cite{Goyal} is used as the benchmark. As for the performance of the communication phase, the fixed weights of the SAC and MO-SAC algorithm dispensing with the meta-framework are invoked as benchmarkers. The parameters of the meta-learning-based MO-SAC network and communication network are summarised at a glance in Table.~\ref{Sim_SAC_network}. The channel coefficients remain the same at each step, but they are different in each episode.

\begin{table*}[tbp]
    \captionsetup[]{font = {small}}
    \caption{Simulation parameters for our meta learning-based MO-SAC algorithm and communication networks \label{Sim_SAC_network}}
    \centering
    \begin{tabular}{lrr}
    \toprule
    Parameter & Description & Value  \\
    \midrule
    $\hat{\mathcal{B}}$ & Mini-batch size for meta-training phase & 64 \\
    $\mathcal{U}_t$ & Number of iterations for meta-training stage & 10000 \\ 
    $\mathcal{D}_t$ & Replay memory for meta-training stage & 1000000 \\
    $\mathcal{U}_{it}$ & Number of iterations for meta-adaptation stage & 10000 \\
    $\mathcal{D}_a$ & Replay memory for meta-adaptation stage  & 1000000 \\
    $(z_V, z_{\pi}, z_{\mathbf{Q}})$ & Inner-loop learning rates & (3, 3, 3) $\times$ $10^{-4}$ \\
    $(\hat{z}_V,\hat{z}_{\pi},\hat{z}_{\mathbf{Q}})$ & Outer-loop learning rates & (3, 3, 3) $\times$ $10^{-3}$\\
    $\mathcal{U}$ & Number of iterations for MO-RL learning phase & 1000 \\
    $\overline{\mathcal{B}}$ & Mini-batch size for MO-RL learning phase & 32 \\
    $\mathcal{D}$ & Replay memory for MO-RL learning phase & 10000 \\
    $\lambda_c$ & MmWave carrier wavelength & 5mm \cite{FTang}\\
    $f$ & THz carrier frequency & 0.2THz \cite{YPan} \\
    $P_{\mathrm{max}}$ & Maximum transmit power & 200mW = 23.01dBm \\
    $h_L$ & Height of LBS & 2m \\
    $h_S$ & Height of CBS & 2.5m \\
    $R_g$ & Length of each grid & 0.05m \\
    $C_{\mathrm{meta}}$ &Cost of MSP& 150 \\
    $t$ &Time slot& 2ms \\
    \bottomrule
    \end{tabular} 
\end{table*}

\vspace{-0.3cm}

\subsection{Performance of Localisation Phase}

\begin{figure}[bp]
    \vspace{-0.8cm}
    \centering
    \subfigure[PEB by Cram\'er-Rao lower bounds.]
    {
        \setlength{\belowcaptionskip}{-0.3cm}
        \begin{minipage}[t]{0.465\textwidth}
            \centering
            \includegraphics[height=2.1in, width=3.2in]{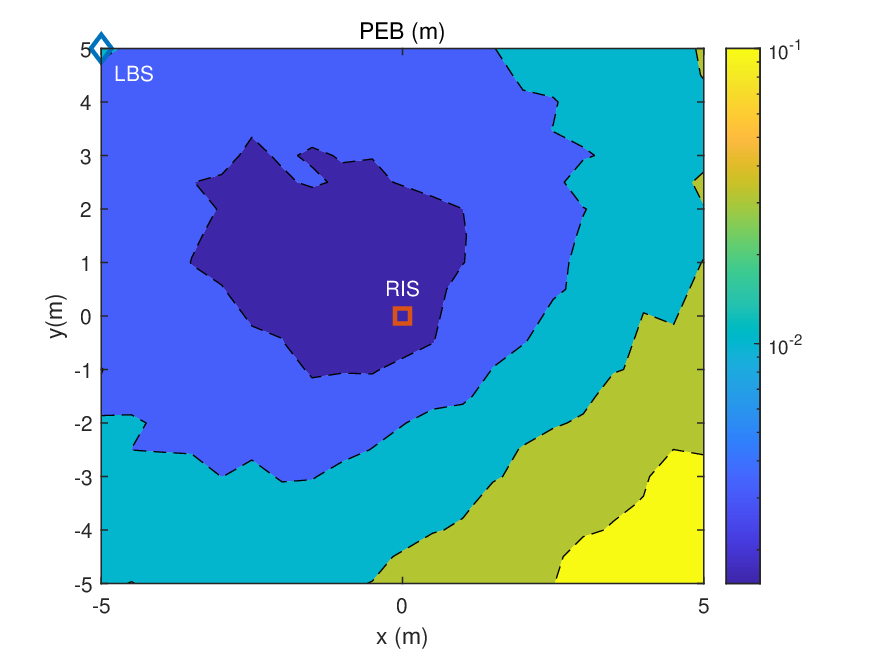}
            \label{PEB}
        \end{minipage}
    }
    \subfigure[PEB by SGD algorithm relying on a minibatch.]
    {   
        \setlength{\belowcaptionskip}{-0.3cm}
        \begin{minipage}[t]{0.465\textwidth}
            \centering
            \includegraphics[height=2.1in, width=3.2in]{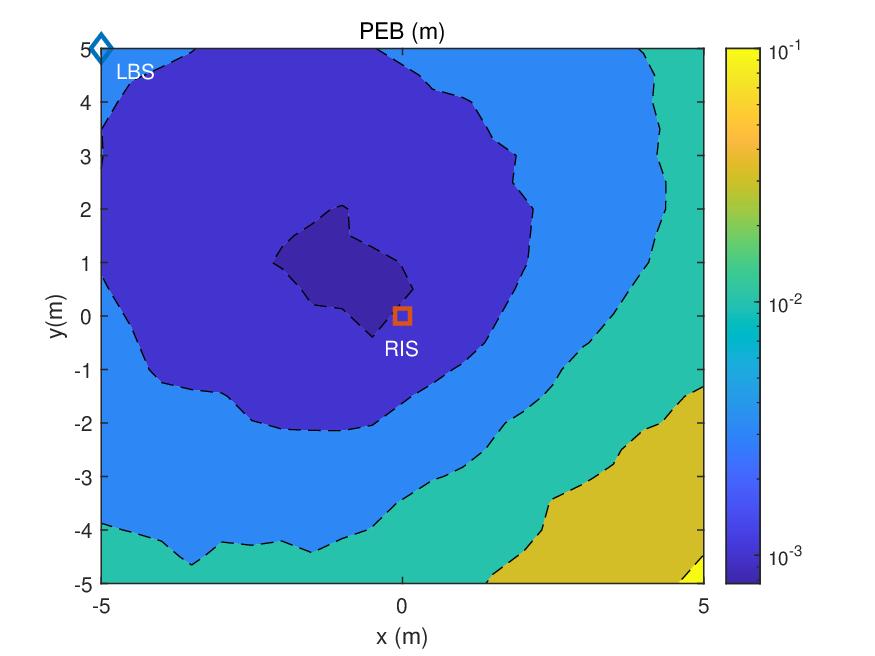}
            \label{PEB_SGD}
        \end{minipage} 
    }
    \caption{PEB by the different algorithms, SNR = 5dB, $K = 128$.}
    \label{PEBall}
\end{figure}

\begin{figure}[tbp]
    \vspace{-0.4cm}
    \setlength{\belowcaptionskip}{-0.5cm}
    \centering
    \subfigure[RMSE of the estimated positions vs the SNR for different algorithms.]
    {
        \begin{minipage}[t]{0.465\textwidth}
            \centering
            \includegraphics[height=2.2in, width=3.2in]{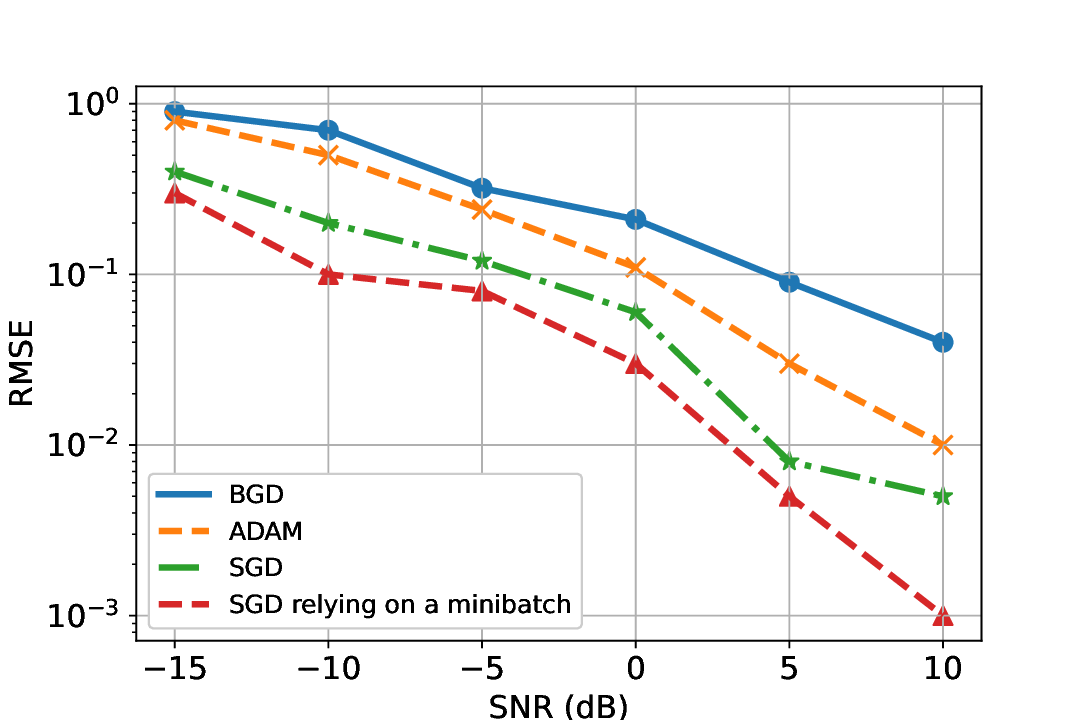}
            \label{p_RMSE}
        \end{minipage}
    }
    \subfigure[RMSE of the estimated positions vs the SNR parameterised by different thresholds $\overline{\eta}$.]
    {
        \begin{minipage}[t]{0.465\textwidth}
            \centering
            \includegraphics[height=2.2in, width=3.2in]{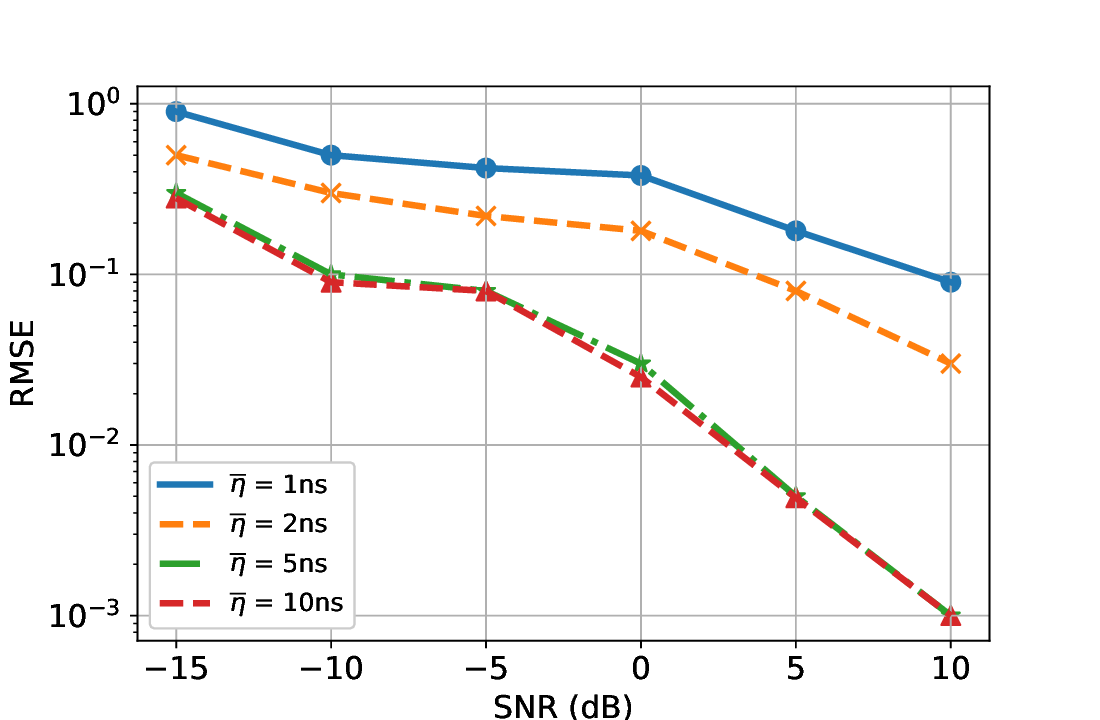}
            \label{eta_RMSE}
        \end{minipage} 
    }
    \caption{RMSE of estimated positions vs the received SNR parameters of Table.~\ref{Sim_SAC_network}.}
    \label{RMSE}
\end{figure}

Fig.~\ref{PEBall} characterises the PEB by the Cram\'er-Rao lower bounds and the SGD algorithm relying on a minibatch, where we have SNR = 5dB, $K = 128$. It is observed that a user position close to the RIS and LBS has a high localisation accuracy. The SGD algorithm relying on a minibatch achieves an excellent accuracy below 10$^{-3}$m, which is more accurate than the PEB based on the Cram\'er-Rao lower bounds. This is because as the distance increases, the signal will also be severely attenuated. Hence the users who are farther away receive poorer signal quality than users who are closer. Additionally, localisation relies on multi-path transmission, and the increase in distance will inevitably increase the noise, which will also affect the localisation accuracy. To further investigate the performance of the proposed SGD algorithm with a minibatch, we compare the root-mean-square error (RMSE) of the estimated positions of different received SNRs and at the different clock offset thresholds shown in Fig.~\ref{RMSE}. Observe in Fig.~\ref{RMSE} that the RMSE is reduced upon increasing the SNR and that compared to the BGD, ADAM, and SGD algorithms, the proposed SGD algorithm relying on a minibatch exhibits better performance. Furthermore, when the SNR achieves 10dB, the RMSE is able to arrive at 10$^{-3}$, and when the maximum tolerable clock offset increases, the RMSE is decreased. This is because the minibatch setting provides the learning experience required, which increases the accuracy of the localisation, and having an increased threshold is capable of increasing the fault tolerance for localisation.

\vspace{-0.3cm}
\subsection{Performance of the Meta-Learning-Based MO-SAC Algorithm}
\begin{figure}[tbp]
    \vspace{-0.4cm}
    \setlength{\belowcaptionskip}{-0.5cm}
    \centering
    \subfigure[Training curves under different algorithms, $K = 128$.]
    {
        \begin{minipage}[t]{0.465\textwidth}
            \centering  
            \includegraphics[height=2.2in, width=3.2in]{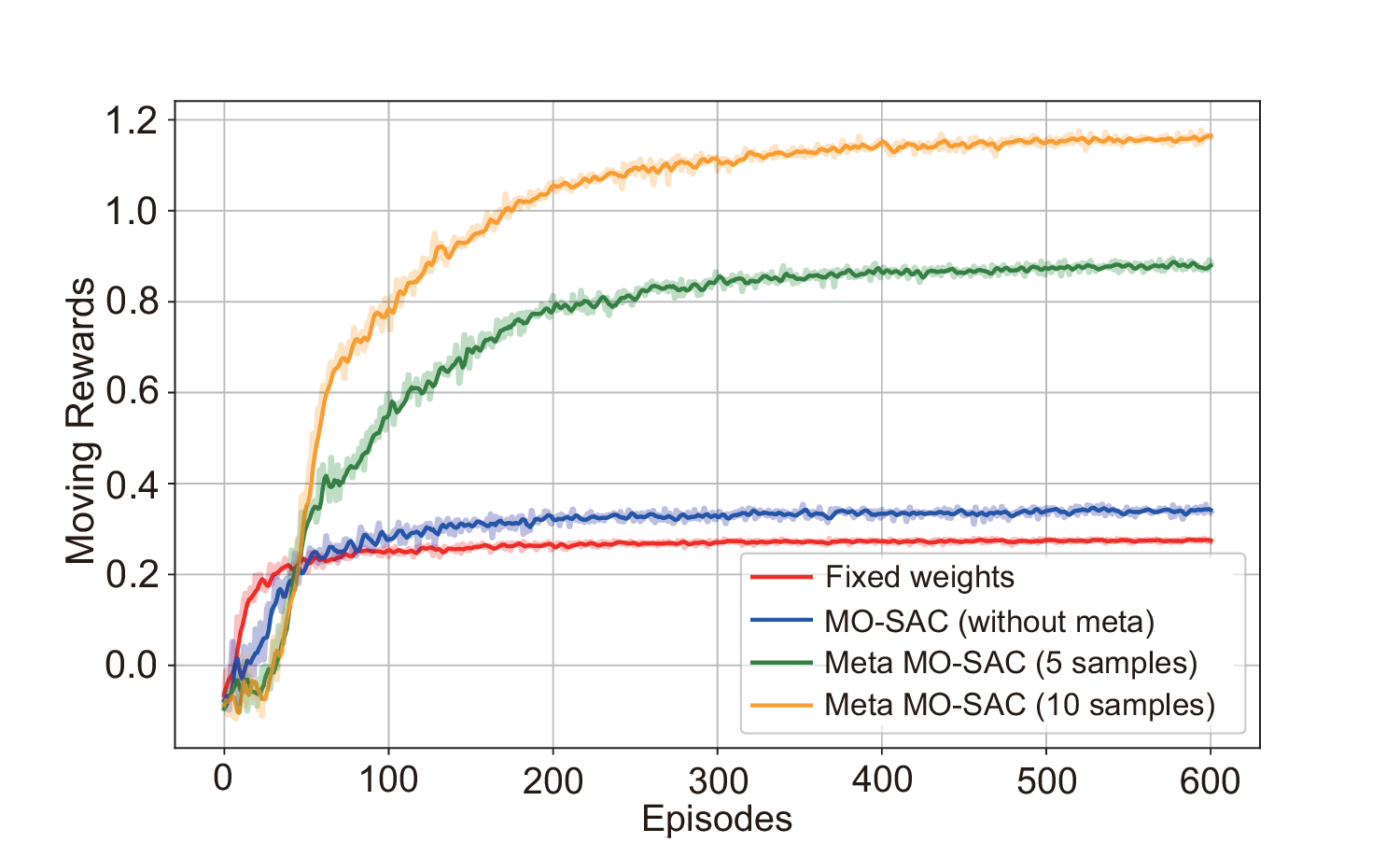}
            \label{Training}
        \end{minipage}
    }
    \subfigure[Adaptation curves under different algorithms, $K = 128$, $N_a = 1$.]
    {
        \begin{minipage}[t]{0.465\textwidth}
            \centering  
            \includegraphics[height=2.2in, width=3.2in]{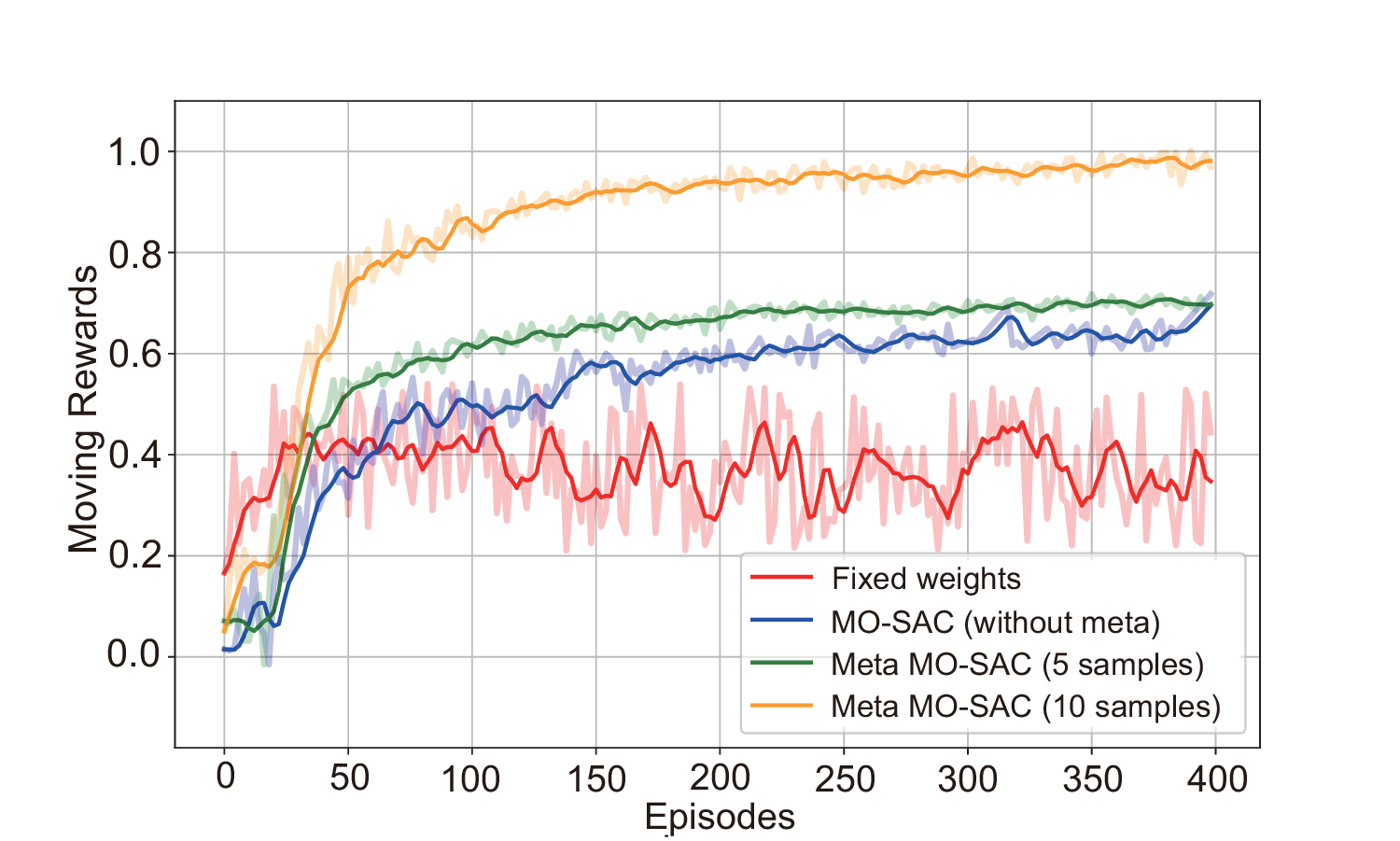}
            \label{Adaptation}
        \end{minipage} 
    }
    \caption{The reward vs episode-index of the proposed algorithm compared to the benchamarkers.}
    \label{performancealgorithm}
\end{figure}

Fig.~\ref{performancealgorithm} characterises the performance of the proposed meta-learning-based MO-SAC algorithm compared to the benchmarkers. Fig.~\ref{Training} illustrates the convergence of the proposed meta-MO-SAC algorithms. The moving reward is the smooth operation for the reward to clearly depict the learning trend of the algorithms. Compared to the benchmarkers, it is observed that the proposed meta-MO-SAC exhibits a slower convergence. This is because of having large quantities of training samples inevitable increases the computational complexity. For the case of fixed weights, although it converges the fastest, it gleans the least reward, so it may not perform as well as the MO algorithm. In the case of the MO-SAC operating without a meta-framework, it converges faster than the meta-learning-based MO-SAC, but its performance has to be further proved according to the adaptation stage. Fig.~\ref{Adaptation} verifies the performance of the trained model when the number of adaptation tasks $N_a$ is set to 1. It is observed that the meta-learning-based MO-SAC algorithm converges promptly to the new task in fewer episodes than the benchmarkers, while the model trained with more training tasks $N$ is capable of exhibiting faster adaption speed to the new tasks. Additionally, the fixed-weights scenario diverges and fails to adapt to new tasks. The MO-SAC algorithm exhibits a slow adaption speed for new tasks, and it is outperformed by the meta-learning-based MO-SAC. Therefore, the adaptation stage illustrates the performance of our proposed meta-algorithm-based MO-SAC algorithms, where the adaption speed can be readily improved upon harnessing longer training.

\vspace{-0.4cm}
\subsection{Approximate PF}

\begin{figure}[tbp]
    \setlength{\belowcaptionskip}{-1cm}
    \centering
    \subfigure[Approximate PF of total service cost vs transmission latency by proposed algorithm compared to the benchmarkers, $K = 128$.]
    {
        \begin{minipage}[t]{0.465\textwidth}
            \centering  
            \includegraphics[height=2.2in, width=3.2in]{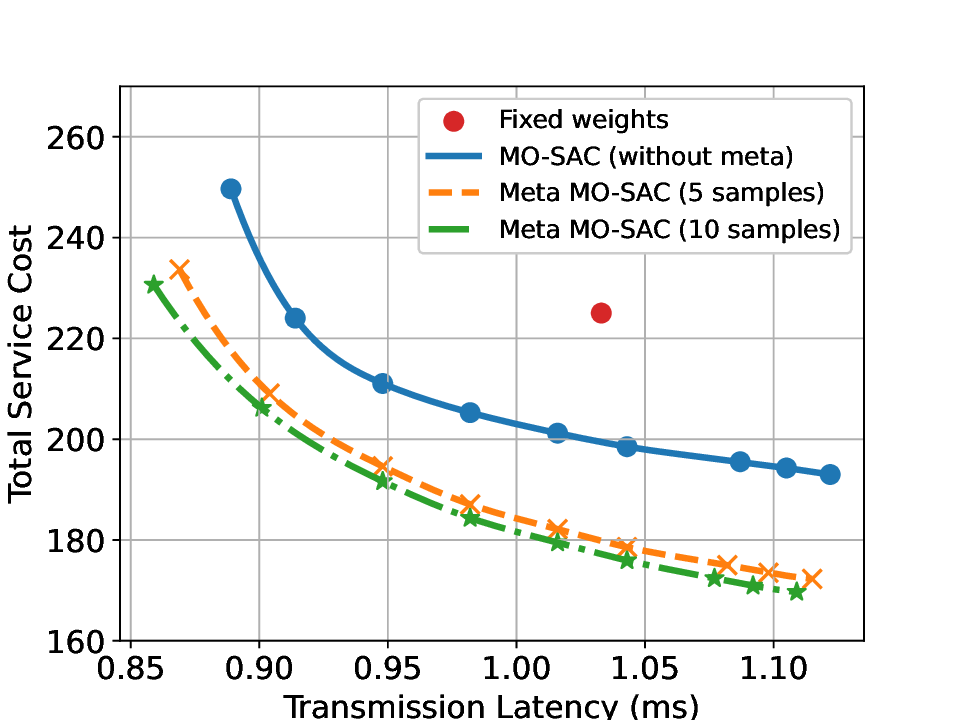}
            \label{PF}
        \end{minipage}
    }
    \subfigure[Approximate PF of Total Service Cost vs Transmission Latency by proposed Meta-MO-SAC algorithm constraint by the different maximum DEP values, $K = 128$.]
    {
        \begin{minipage}[t]{0.465\textwidth}
            \centering  
            \includegraphics[height=2.2in, width=3.2in]{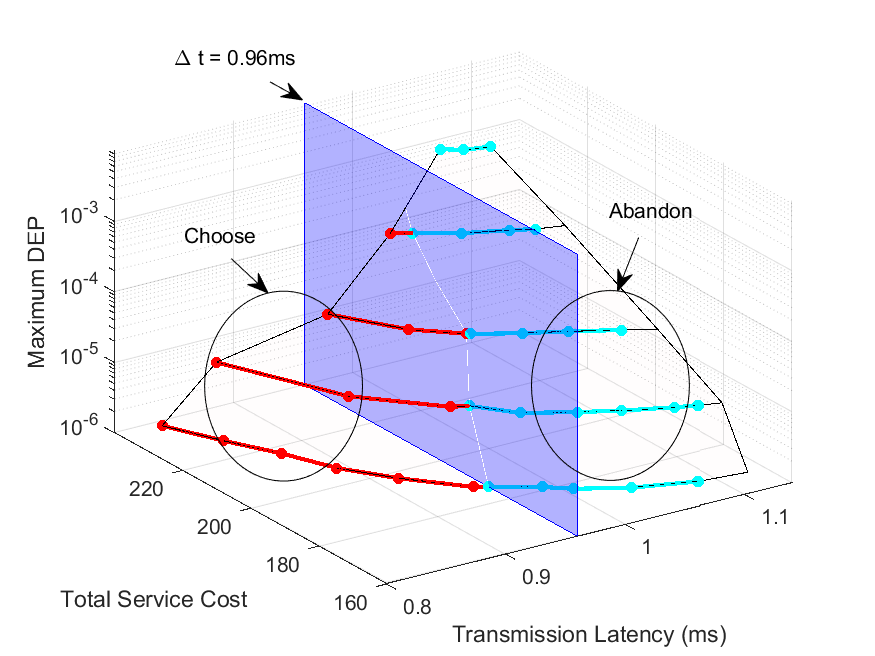}
            \label{PF_3D}
        \end{minipage} 
    }
    \caption{The Approximate PF of Total Service Cost vs Transmission Latency by the proposed algorithm compared to the benchamarkers.}
    \label{pareto}
\end{figure}

Fig.~\ref{pareto} depicts the approximate PFs based on the adaptation stage, since it has constrained options. As mentioned in the formulated problem, our goal is to strike a transmission latency versus total service cost tradeoff. As shown in Fig.~\ref{PF}, the approximate PF found by the proposed meta-MO-SAC algorithm approaches the coordinates, which covers more feasible solutions than the benchmarkers. Additionally, the model trained based on numerous training samples outperforms that based on a few training samples. As shown in Fig.~\ref{PF_3D}, the approximate PF changes with maximum DEP constraint. With the increase of the maximum DEP, the coverage of PF shrinks, which indicates that the DEP plays an important role in determining the coverage of PF. This is because when the maximum DEP decreases, the achievable data rate range becomes wider and further illustrates the low latency achieved for the entire system. Then, according to Equations \eqref{reliability state} - \eqref{reliability}, we have set a maximum tolerable transmission delay for the entire system to characterise the requirements of a practical scenario. For example, when the maximum tolerable transmission delay $\Delta t$ is set to 0.96ms, the right-hand side of the approximate PF in Fig.~\ref{PF_3D} will be abandoned and the transmission latency below 0.96ms is chosen. Furthermore, the trend is that the PF becomes shorter with the increase of the maximum DEP and there is no obvious tradeoff between the total service cost and the transmission latency. Hence, the associated MO optimization can be formulated as a single objective optimization, when the maximum DEP is high enough. This insight provides a guideline for Metaverse deployments by MSP and NSP.


\vspace{-0.3cm}

\subsection{Different Number of Elements in the RIS}
\begin{figure}[htbp]
    \vspace{-0.3cm}
    \setlength{\belowcaptionskip}{-0.6cm}
    \centering
    \includegraphics[height=2.2in, width=3.2in]{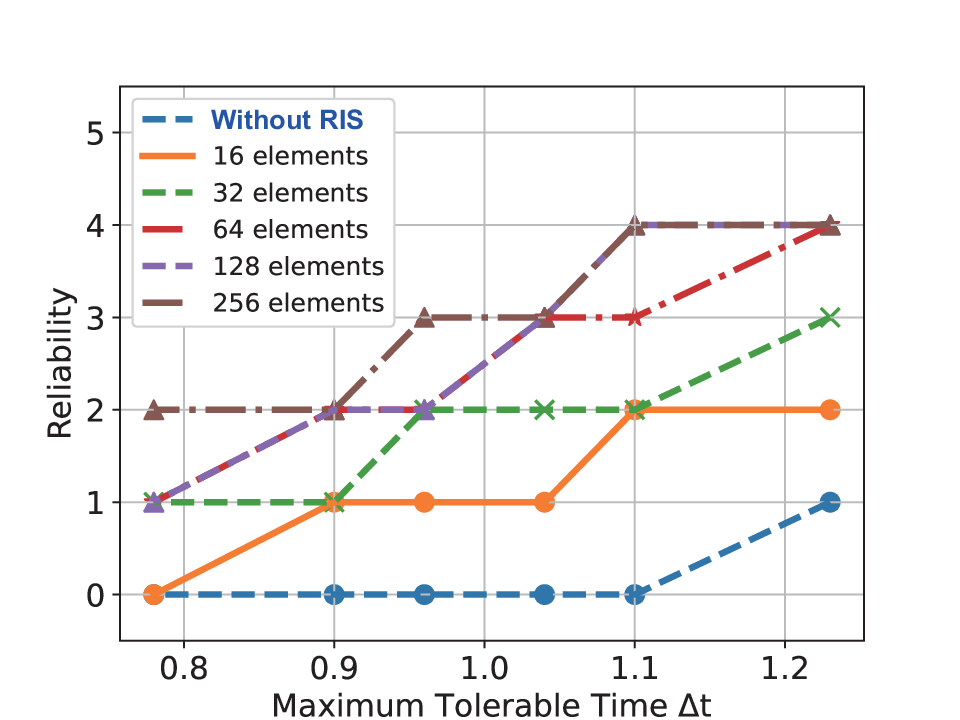}
    \caption{Reliability vesus Maximum Latency $\Delta t$, DEP = $10^{-5}$, $N_a = 1$.}
    \label{ele}
\end{figure}

Fig.~\ref{ele} quantifies the impact of the different number of RIS elements on the system. According to the settings of $\Delta t$ in Fig.~\ref{PF_3D}, when the maximum tolerable transmission latency increases, the range of PF increases. Having a high latency tolerable increases the probability of users being served successfully. Therefore, the number of users served successfully increases, as the maximum tolerable transmission latency increases. In this case, when the number of elements in the RIS increases, the reliability is improved, because the channel gains of a communication system are improved. Therefore, these results demonstrate that RISs have a positive impact on a communication system.

\vspace{-0.3cm}
\section{Conclusions}
Two KPIs with conflicting relationships, i.e., total service cost and transmission latency, were considered in URLLC-enabled Metaverse. Our designed objective was to simultaneously minimise the total service cost and transmission latency in URLLC-based Metaverse networks approaching PF, while jointly optimising the transmit power, the RIS phase shifts, and the decoding error probability. To solve this problem, a meta-learning-based position-dependent MO-SAC algorithm was proposed. The core idea of the proposed algorithm was to dynamically assign the weights each time, when a network update process takes place. Additionally, there were two key points to assist the algorithm: the SGD algorithm with a mini-batch was harnessed for determining the positions of the users, while the MO-SAC algorithm was invoked for learning a policy for each task. The numerical results showed that: 1) The proposed solution struck a compelling tradeoff between the total service cost and transmission latency, which provided a candidate group of optimal solutions for practical scenarios. 2) The proposed meta-learning-based MO-SAC algorithm was capable of achieving a faster adaptation to new wireless environments than the benchmarkers. 3) The approximate PF discovered the relationships among the KPIs for the Metaverse, which provided guidelines for its deployment. Furthermore, extending our ML algorithm to the general system studied is the main focus in the next step of research.

\appendices
\vspace{-0.3cm}
\section*{Appendix~A: Proof of Theorem \ref{theorem 1}} \label{Appendix:A.1}
\renewcommand{\theequation}{A.\arabic{equation}}
\setcounter{equation}{0}
\setcounter{theorem}{0}

\begin{theorem}\label{theorem 1p}
    Let $\mathbf{Q}$' be the preferred optimal value function formulated as follows:
    \onecolumn
    \par
    \vspace{-0.1cm}
    \noindent
    \begin{align}\label{pargQp1}
        \mathbf{Q}'(\overline{\mathbf{s}}_t, \overline{\mathbf{a}}_t, \mathbf{w}) = \Big\{\mathrm{arg}_{\mathbf{Q}} \mathrm{sup}_{\pi \in \Pi} \mathbf{w}^H \mathbb{E}_{\overline{\mathbf{s}}_{t+1} \sim \zeta_{\pi}}  \Big[\sum_{t=0}^{T} \gamma_0^{t} \overline{\mathbf{r}}(\overline{\mathbf{s}}_t, \overline{\mathbf{a}}_t, \mathbf{w})\Big] - \mathbf{w}^{'H}\mathrm{log}\pi(\overline{\mathbf{a}}_t|\overline{\mathbf{s}}_t, \mathbf{w})\Big\}.
    \end{align}
    \par
    \vspace{-0.1cm}
    \noindent
    Then, it can be obtained that: $\mathbf{V}'(\overline{\mathbf{s}}_t, \overline{\mathbf{a}}_t, \mathbf{w}) = \mathcal{G} \mathbf{Q}'(\overline{\mathbf{s}}_t, \overline{\mathbf{a}}_t, \mathbf{w})$.
    \par
    \vspace{-0.2cm}
    \noindent
    \begin{proof}
        Firstly, upon taking into account that $\overline{d}(\mathbf{Q}', \mathcal{G} \mathbf{V}') = \mathrm{sup}_{\overline{\mathbf{a}}_t \in \overline{\mathbf{A}}, \mathbf{w} \subseteq \mathbf{W}} |\mathbf{w}^H \mathbf{Q}(\overline{\mathbf{s}}_t, \overline{\mathbf{a}}_t, \mathbf{w}) - \mathcal{G} \mathbf{V}(\overline{\mathbf{s}}_t, \overline{\mathbf{a}}_t, \mathbf{w})| = 0$, we are able to observe that $ \mathbf{w}^H \mathcal{G} \mathbf{Q}(\overline{\mathbf{s}}_t, \overline{\mathbf{a}}_t, \mathbf{w}) = \mathbf{w}^H \mathbf{V}(\overline{\mathbf{s}}_t, \overline{\mathbf{a}}_t, \mathbf{w})$. The proof is provided as follows:
        \par
        \vspace{-0.1cm}
        \noindent
        \begin{align}\label{proof1}
            &\mathbf{w}^H \mathcal{G} \mathbf{Q}(\overline{\mathbf{s}}_t, \overline{\mathbf{a}}_t, \mathbf{w}) = \mathbf{w}^H \overline{\mathbf{r}}(\overline{\mathbf{s}}_t, \overline{\mathbf{a}}_t, \mathbf{w}) + \mathbf{w}^H \gamma_0\mathbb{E}_{\overline{\mathbf{s}}_{t+1} \sim \zeta_{\pi}}[V(\overline{\mathbf{s}}_{t+1}, \mathbf{w})], \nonumber \\
            &= \mathbf{w}^H \overline{\mathbf{r}}(\overline{\mathbf{s}}_t, \overline{\mathbf{a}}_t, \mathbf{w}) + \mathbf{w}^H \gamma_0\mathbb{E}_{\overline{\mathbf{s}}_{t+1} \sim \zeta_{\pi}} \arg_Q \mathrm{sup}_{\overline{\mathbf{a}}_t \in \overline{\mathbf{A}}, \mathbf{w}' \subseteq \mathbf{W}} [\mathbf{w}^H \mathbf{Q}(\overline{\mathbf{s}}_t, \overline{\mathbf{a}}_t,\mathbf{w}') - \mathbf{w}^H\mathrm{log}\pi(\overline{\mathbf{a}}_t|\overline{\mathbf{s}}_t, \mathbf{w}')], \nonumber \\
            &= \mathbf{w}^H \overline{\mathbf{r}}(\overline{\mathbf{s}}_t, \overline{\mathbf{a}}_t, \mathbf{w}) + \gamma_0\mathbb{E}_{\overline{\mathbf{s}}_{t+1} \sim \zeta_{\pi}} \mathrm{sup}_{\overline{\mathbf{a}}_t \in \overline{\mathbf{A}}, \mathbf{w}' \subseteq \mathbf{W}} [\mathbf{w}^H \mathbf{Q}(\overline{\mathbf{s}}_t, \overline{\mathbf{a}}_t,\mathbf{w}') - \mathbf{w}^H\mathrm{log}\pi(\overline{\mathbf{a}}_t|\overline{\mathbf{s}}_t, \mathbf{w}')], \nonumber \\
            &= \mathbf{w}^H \overline{\mathbf{r}}(\overline{\mathbf{s}}_t, \overline{\mathbf{a}}_t, \mathbf{w}) + \gamma_0\mathbb{E}_{\overline{\mathbf{s}}_{t+1} \sim \zeta_{\pi}} \mathrm{sup}_{\overline{\mathbf{a}}_t \in \overline{\mathbf{A}}, \mathbf{w}' \subseteq \mathbf{W}} \Big\{\mathbf{w}^H \Big\{\mathrm{arg}_{\mathbf{Q}} \mathrm{sup}_{\pi \in \Pi} \mathbf{w}^{'H} \mathbb{E}_{\overline{\mathbf{s}}_{t+1} \sim \zeta_{\pi}}  \Big[\sum_{t=0}^{T} \gamma_0^{t} \overline{\mathbf{r}}(\overline{\mathbf{s}}_t, \overline{\mathbf{a}}_t, \mathbf{w})\Big] \nonumber \\
            & \hspace{1em} - \mathbf{w}^{'H}\mathrm{log}\pi(\overline{\mathbf{a}}_t|\overline{\mathbf{s}}_t, \mathbf{w})\Big\}\Big\}, \nonumber \\
            &= \mathbf{w}^H \overline{\mathbf{r}}(\overline{\mathbf{s}}_t, \overline{\mathbf{a}}_t, \mathbf{w}) + \gamma_0\mathbb{E}_{\overline{\mathbf{s}}_{t+1} \sim \zeta_{\pi}} \mathrm{sup}_{\overline{\mathbf{a}}_t \in \overline{\mathbf{A}}} \Big\{\mathbf{w}^H \Big\{\mathrm{arg}_{\mathbf{Q}} \mathrm{sup}_{\pi \in \Pi} \mathbb{E}_{\overline{\mathbf{s}}_{t+1} \sim \zeta_{\pi}}  \Big[\sum_{t=0}^{T} \gamma_0^{t} \overline{\mathbf{r}}(\overline{\mathbf{s}}_t, \overline{\mathbf{a}}_t, \mathbf{w})\Big]\Big\}\Big\}, \nonumber \\
            &= \mathbf{w}^H \Big\{\mathrm{arg}_{\mathbf{Q}} \mathrm{sup}_{\pi \in \Pi} \mathbf{w}^H \Big\{\mathbb{E}_{\overline{\mathbf{s}}_{t+1} \sim \zeta_{\pi}}  \Big[\sum_{t=0}^{T} \gamma_0^{t} \overline{\mathbf{r}}(\overline{\mathbf{s}}_t, \overline{\mathbf{a}}_t, \mathbf{w})\Big] - \mathrm{log}\pi(\overline{\mathbf{a}}_t|\overline{\mathbf{s}}_t, \mathbf{w})\Big\}\Big\}, \nonumber \\
            &= \mathbf{w}^H \mathbf{V}(\overline{\mathbf{s}}_t, \overline{\mathbf{a}}_t, \mathbf{w}).
        \end{align}
        \par
        \vspace{-0.1cm}
        \noindent
        Since we have the $\overline{d}(\mathbf{Q}', \mathcal{G} \mathbf{V}') = 0$, the preferred optimal value function is a fixed point of the proposed optimality operator. This concludes the proof of \textbf{Theorem \ref{theorem 1p}}.
    \end{proof}
\end{theorem}

\vspace{-0.5cm}
\section*{Appendix~B: Proof of Theorem \ref{theorem 2}} \label{Appendix:A.2}
\renewcommand{\theequation}{B.\arabic{equation}}
\setcounter{equation}{0}

\begin{theorem}\label{theorem 2p}
    Let us define a pair of MO Q value functions by $\mathbf{Q}$ and $\hat{\mathbf{Q}}$. Then the Lipschitz condition of $\overline{d}(\mathcal{G}\mathbf{Q}, \mathcal{G}\hat{\mathbf{Q}}) \leq \gamma_0 \overline{d}(\mathbf{Q}, \hat{\mathbf{Q}})$ can be satisfied, where $\gamma_0$ is the discount factor.
    \begin{proof}
    Upon assuming $\mathrm{sup}_{\overline{\mathbf{a}}_t \in \overline{\mathbf{A}}, \mathbf{w}' \subseteq \mathbf{W}} \mathbf{w}^H \mathbf{Q}(\overline{\mathbf{s}}_t, \overline{\mathbf{a}}_t, \mathbf{w}') \geq \mathrm{sup}_{\overline{\mathbf{a}}_t \in \overline{\mathbf{A}}, \mathbf{w}' \subseteq \mathbf{W}} \mathbf{w}^H \hat{\mathbf{Q}}(\overline{\mathbf{s}}_t, \overline{\mathbf{a}}_t, \mathbf{w}')$, we have
    \par
    \vspace{-0.2cm}
    \noindent
    \begin{align}\label{proof2}
        &\overline{d}(\mathcal{G}\mathbf{Q}, \mathcal{G}\hat{\mathbf{Q}}) = \mathrm{sup}_{\overline{\mathbf{s}}_t \in \overline{\mathbf{S}}, \overline{\mathbf{a}}_t \in \overline{\mathbf{A}}, \mathbf{w}' \subseteq \mathbf{W}} |\mathbf{w}^H (\mathbf{Q}(\overline{\mathbf{s}}_t, \overline{\mathbf{a}}_t, \mathbf{w}') - \hat{\mathbf{Q}}(\overline{\mathbf{s}}_t, \overline{\mathbf{a}}_t, \mathbf{w}'))| \nonumber \\
        &=\mathrm{sup}_{\overline{\mathbf{s}}_t \in \overline{\mathbf{S}}, \overline{\mathbf{a}}_t \in \overline{\mathbf{A}}, \mathbf{w}' \subseteq \mathbf{W}} |\gamma_0 \mathbf{w}^H \mathbb{E}_{\overline{\mathbf{s}}_{t+1} \sim \zeta_{\pi}}  [\mathbf{Q}(\overline{\mathbf{s}}_t, \overline{\mathbf{a}}_t, \mathbf{w})] - \gamma_0 \mathbf{w}^H \mathbb{E}_{\overline{\mathbf{s}}_{t+1} \sim \zeta_{\pi}}  [\hat{\mathbf{Q}}(\overline{\mathbf{s}}_t, \overline{\mathbf{a}}_t, \mathbf{w})]| \nonumber \\
        &\leq \gamma_0 \mathrm{sup}_{\overline{\mathbf{s}}_t \in \overline{\mathbf{S}}, \mathbf{w}' \subseteq \mathbf{W}}|\mathbf{w}^H [\mathrm{arg}_Q \mathrm{sup}_{\overline{\mathbf{a}}_t \in \overline{\mathbf{A}}, \mathbf{w}' \subseteq \mathbf{W}}\mathbf{Q}(\overline{\mathbf{s}}_t, \overline{\mathbf{a}}_t, \mathbf{w}')- \mathrm{arg}_Q \mathrm{sup}_{\overline{\mathbf{a}}'_t \in \overline{\mathbf{A}}, \mathbf{w}'' \subseteq \mathbf{W}}\hat{\mathbf{Q}}(\overline{\mathbf{s}}_t, \overline{\mathbf{a}}_t, \mathbf{w}'')]|, \nonumber \\
        &\leq \gamma_0 \mathrm{sup}_{\overline{\mathbf{s}}_t \in \overline{\mathbf{S}}, \mathbf{w}' \subseteq \mathbf{W}}|\mathrm{sup}_{\overline{\mathbf{a}}_t \in \overline{\mathbf{A}}, \mathbf{w}' \subseteq \mathbf{W}}\mathbf{w}^H\mathbf{Q}(\overline{\mathbf{s}}_t, \overline{\mathbf{a}}_t, \mathbf{w}')- \mathrm{sup}_{\overline{\mathbf{a}}'_t \in \overline{\mathbf{A}}, \mathbf{w}'' \subseteq \mathbf{W}}\mathbf{w}^H\hat{\mathbf{Q}}(\overline{\mathbf{s}}_t, \overline{\mathbf{a}}_t, \mathbf{w}'')|, \nonumber \\
        &= \gamma_0 \mathrm{sup}_{\overline{\mathbf{s}}_t \in \overline{\mathbf{S}}, \mathbf{w}' \subseteq \mathbf{W}}|\mathbf{w}^H\mathbf{Q}(\overline{\mathbf{s}}_t, \overline{\mathbf{a}}_t, \mathbf{w}') - \mathbf{w}^H\hat{\mathbf{Q}}(\overline{\mathbf{s}}_t, \overline{\mathbf{a}}_t, \mathbf{w}') + \mathbf{w}^H\hat{\mathbf{Q}}(\overline{\mathbf{s}}_t, \overline{\mathbf{a}}_t, \mathbf{w}') - \mathrm{sup}_{\overline{\mathbf{a}}'_t \in \overline{\mathbf{A}}, \mathbf{w}'' \subseteq \mathbf{W}}\mathbf{w}^H\hat{\mathbf{Q}}(\overline{\mathbf{s}}_t, \overline{\mathbf{a}}_t, \mathbf{w}'')|, \nonumber \\
        &\leq \gamma_0 \mathrm{sup}_{\overline{\mathbf{s}}_t \in \overline{\mathbf{S}}, \mathbf{w}' \subseteq \mathbf{W}}|\mathbf{w}^H\mathbf{Q}(\overline{\mathbf{s}}_t, \overline{\mathbf{a}}_t, \mathbf{w}') - \mathbf{w}^H\hat{\mathbf{Q}}(\overline{\mathbf{s}}_t, \overline{\mathbf{a}}_t, \mathbf{w}')|, \nonumber \\
        &\leq \gamma_0 \mathrm{sup}_{\overline{\mathbf{s}}_t \in \overline{\mathbf{S}}, \mathbf{a}_t \in \overline{\mathbf{A}}, \mathbf{w}' \subseteq \mathbf{W}}|\mathbf{w}^H\mathbf{Q}(\overline{\mathbf{s}}_t, \overline{\mathbf{a}}_t, \mathbf{w}') - \mathbf{w}^H\hat{\mathbf{Q}}(\overline{\mathbf{s}}_t, \overline{\mathbf{a}}_t, \mathbf{w}')|, \nonumber \\
        &= \gamma_0 \overline{d}(\mathbf{Q}, \hat{\mathbf{Q}}).
    \end{align}
    \par
    \vspace{-0.1cm}
    \noindent
    Then, we proved \textbf{Theorem \ref{theorem 2p}}.
    \end{proof}
\end{theorem}

\vspace{-0.8cm}
\section*{Appendix~C: Proof of Theorem \ref{theorem 3}} \label{Appendix:A.3}
\renewcommand{\theequation}{C.\arabic{equation}}
\setcounter{equation}{0}

\begin{theorem}\label{theorem 3p}
    If $\mathcal{G}$ can be contracted with the aid of the discount factor $\gamma_0$ on the complete pseudo-metric space $\langle\pmb{\mathcal{Q}}, \overline{d}\rangle$, it can be shown that $\mathrm{lim}_{t \rightarrow \infty} \overline{d}(\mathcal{G}^{t}, \mathbf{Q}, \hat{\mathbf{Q}}) = 0$, where $\pmb{\mathcal{Q}}$ is the value space.
    \begin{proof}
    Similarly to \textbf{Theorem \ref{theorem 2}}, we define a pair of mo Q value functions by $\mathbf{Q}$ and $\hat{\mathbf{Q}}$. we have:
    \par
    \vspace{-0.2cm}
    \noindent
    \begin{align}\label{proof3}
        \overline{d}(\mathbf{Q}, \hat{\mathbf{Q}}) \leq \overline{d}(\mathbf{Q}, \mathcal{G}\mathbf{Q}) + \overline{d}(\mathcal{G}\mathbf{Q}, \mathcal{G}\hat{\mathbf{Q}}) + \overline{d}(\mathcal{G}\hat{\mathbf{Q}}, \hat{\mathbf{Q}})\leq\overline{d}(\mathbf{Q}, \mathcal{G}\mathbf{Q}) + \gamma_0 \overline{d}(\mathbf{Q}, \hat{\mathbf{Q}}) + \overline{d}(\mathcal{G}\hat{\mathbf{Q}}, \hat{\mathbf{Q}}),
    \end{align}
    \par
    \vspace{-0.1cm}
    \noindent
    and it can be shown that:
    \par
    \vspace{-0.3cm}
    \noindent
    \begin{align}
        \overline{d}(\mathbf{Q}, \hat{\mathbf{Q}}) \leq \frac{\overline{d}(\mathbf{Q}, \mathcal{G}\mathbf{Q}) + \overline{d}(\hat{\mathbf{Q}}, \mathcal{G}\hat{\mathbf{Q}})}{1-\gamma_0}.
    \end{align}
    \par
    \vspace{-0.1cm}
    \noindent
    Upon considering the pair of values $\mathcal{G}^{t_1}\mathbf{Q}$ and $\mathcal{G}^{t_2}\mathbf{Q}$ in \{$\mathcal{G}^{t}\mathbf{Q}$\}, the following upper bound can be formulated:
    \par
    \vspace{-0.1cm}
    \noindent
    \begin{align}
        \overline{d}(\mathcal{G}^{t_1}\mathbf{Q}, \mathcal{G}^{t_2}\mathbf{Q}) &\leq \frac{\overline{d}(\mathcal{G}^{t_1}\mathbf{Q}, \mathcal{G}^{t_1+1}\mathbf{Q}) + \overline{d}(\mathcal{G}^{t_2}\hat{\mathbf{Q}}, \mathcal{G}^{t_2+1}\hat{\mathbf{Q}})}{(1-\gamma_0)} \leq \frac{\gamma_{t_1}\overline{d}(\mathbf{Q}, \mathcal{G}\mathbf{Q}) + \gamma_{t_2}\overline{d}(\mathbf{Q}, \mathcal{G}\mathbf{Q})}{1-\gamma_0} \leq \frac{\gamma_{t_1}+\gamma_{t_2}}{1-\gamma_0}\overline{d}(\mathbf{Q}, \mathcal{G}\mathbf{Q}).
    \end{align}
    \par
    \vspace{-0.1cm}
    \noindent
    Since $\gamma_0 \in [0,1)$, for any two values of $\mathcal{G}^{t_1}\mathbf{Q}$, $\mathcal{G}^{t_2}\mathbf{Q}$, we can have:
    \par
    \vspace{-0.1cm}
    \noindent
    \begin{align}
        \overline{d}(\mathcal{G}^{t_1}\mathbf{Q}, \mathbf{Q}) = \lim_{t_1 \leftarrow \infty} \overline{d}(\mathcal{G}^{t_1+1}\mathbf{Q}, \mathbf{Q})  = \lim_{t_1 \leftarrow \infty} \overline{d}(\mathcal{G}^{t_1}\mathbf{Q}, \mathbf{Q}) = 0, \nonumber \\
        \overline{d}(\mathcal{G}^{t_2}\mathbf{Q}, \mathbf{Q}) = \lim_{t_2 \leftarrow \infty} \overline{d}(\mathcal{G}^{t_2+1}\mathbf{Q}, \mathbf{Q})  = \lim_{t_2 \leftarrow \infty} \overline{d}(\mathcal{G}^{t_2}\mathbf{Q}, \mathbf{Q}) = 0.
    \end{align}
    \par
    \vspace{-0.1cm}
    \noindent
    Hence, we proved \textbf{Theorem \ref{theorem 3p}}. 
    \end{proof}
\end{theorem}

\begin{multicols}{2}

\end{multicols}
\end{sloppypar}
\end{document}